\documentclass[journal]{IEEEtran}
\usepackage{amsmath,amsfonts}
\usepackage{algorithmic}
\usepackage{algorithm}
\usepackage{array}
\usepackage[caption=false,font=normalsize,labelfont=sf,textfont=sf]{subfig}
\usepackage{textcomp}
\usepackage{stfloats}
\usepackage{url}
\usepackage{verbatim}
\usepackage{graphicx}
\usepackage{cite}
\usepackage{pslatex}
\usepackage{balance}
\usepackage{booktabs}
\usepackage{xcolor}
\usepackage{makecell}

\newcommand{\aname}{Accel-Align}
\newcommand{\labone}{Exp. 1}
\newcommand{\labtwo}{Exp. 2}
\newcommand{\labthree}{Exp. 3}

\usepackage{etoolbox}
\makeatletter
\patchcmd{\@makecaption}
  {\scshape}
  {}
  {}
  {}
\patchcmd{\@makecaption}
  {\\}
  {.\ }
  {}
  {}  
\makeatother

\begin{document}

\title{CMOSS: A Reliable, Motif-based Columnar Molecular Storage System}
\author{Eugenio Marinelli$^*$, Yiqing Yan$^*$, Virginie Magnone,  Pascal Barbry, Raja Appuswamy
\IEEEcompsocitemizethanks{\IEEEcompsocthanksitem Eugenio Marinelli, Yiqing Yan and Raja Appuswamy are with EURECOM, France.
\IEEEcompsocthanksitem Virginie Magnone and Pascal Barbry are with IPCM - Institut de Pharmacologie Moléculaire et Cellulaire, France.}%
}%


\maketitle

\begin{abstract}

The surge in demand for cost-effective, durable long-term archival media, coupled with density limitations of contemporary magnetic media, has resulted in synthetic DNA emerging as a promising new alternative. Despite its benefits, storing data on DNA poses several challenges as the technology used for reading/writing data and achieving random access on DNA are highly error prone. 

In order to deal with such errors, it is important to design efficient pipelines that can carefully use redundancy to mask errors without amplifying overall cost. In this work, we present Columnar MOlecular Storage System (CMOSS), a novel, end-to-end DNA storage pipeline that can provide error-tolerant data storage at low read/write costs. CMOSS differs from SOTA on three fronts (i) a motif-based, vertical layout in contrast to nucleotide-based horizontal layout used by SOTA, (ii) merged consensus calling and decoding enabled by the vertical layout, and (iii) a flexible, fixed-size, block-based data organization for random access over DNA storage in contrast to the variable-sized, object-based access used by SOTA. Using an in-depth evaluation via simulation studies and real wet-lab experiments, we demonstrate the benefits of various CMOSS design choices.
We make the entire pipeline together with the read datasets openly available to the community for faithful reproduction and furthering research.





\end{abstract}


\section{Introduction}


The global datasphere, or the sum total of all digital data generated, is expected to reach 125 Zettabytes by 2025, and over 50\% of such data will be enterprise data stored in various databases, data lakes, and warehouses~\cite{IDCReport}. Today, over 80\% of data generated is ``cold'', or infrequently accessed, and corresponds to data that needs to be archived in order to meet safety, legal and regulatory compliance requirements~\cite{IntelWhitePaper}. Archival data is the fastest growing data segment with over 60\% cumulative annual growth rate~\cite{HorisonReport}. As enterprises databases continue migrate to the cloud, cloud vendors are in need of archival storage technologies that can provide high-density, low-cost storage of such data for decades without degradation. As all current storage media suffer from density scaling and durability limitations~\cite{semiconductor-synbio,oligoarchive}, researchers have started investigating radically new medium optimized for long-term archival. One such medium that has received a lot of attention recently is synthetic DNA.

DNA as a storage medium is seven orders of magnitude denser than tape~\cite{semiconductor-synbio} and can store up to 1 Exabyte of data in a cubic millimeter~\cite{church-synthesis}. It is extremely durable and can last several millenia when stored under proper conditions. DNA is read by a process called sequencing, and the sequencing technology used to DNA is decoupled from DNA, the storage medium, itself. Thus, DNA will not suffer from obsolescence issues as we will always be able to read back data stored in DNA. Finally, using common, well-established biochemical techniques, it is very easy to replicate DNA rapidly. Thus, data stored in DNA can be easily copied. Given these benefits, several researchers have demonstrated the feasibility of using DNA as a long-term archival storage medium~\cite{church-synthesis,Blawat-fec,erlich2017dna,organick2018random,goldman2013towards,grass2015robust,ldpc-dna,oligoarchive,dnaskew-isca22}.

The biochemical processes used for writing (\emph{synthesis}) and reading (\emph{sequencing}) DNA are not precise operations as they introduce a variety of errors. In order to provide reliable data storage on DNA despite such errors, it is necessary to use redundancy in both writing (using some form of error control coding) and reading pipelines (in the form of very high sequencing coverage). The added redundancy has the undesirable side effect of amplifying the read/write cost. Thus, efficient handling of errors is crucial to reducing overall cost. Similarly, given the high density of DNA, an archive stored in DNA can contain millions to billions of files. However, real-world scenarios often demand access to only a fraction of the information. For instance, clients may request a single table from a database or extract a specific image from a collection. Sequencing the entire stored information in response to these requests is needless, expensive, and time consuming, as decoding the information stored in DNA-based storage involves applying a computationally-intensive consensus calling procedure to aggregate reads originating from the same oligo. Thus, the implementation of reliable random access is crucial in making large-scale, cost-efficient DNA-based storage feasible.

In this work, we present \emph{Columnar MOlecular Storage Systems (CMOSS)}, an end-to-end pipeline for DNA storage~\footnote{Source code: https://gitlab.eurecom.fr/marinele/oligoarchive-columnar.git, data will be made available in this repository once we make it accessible} that provides substantially lower read/write costs than SOTA approaches. Our work is orthogonal to the growing literature on designing optimal error-correction codes for DNA storage in that CMOSS can be used with any error-control code. The key aspects of CMOSS that distinguish it are: (i) a novel, motif-based, vertical, cross-oligo layout for DNA storage in contrast to the nucleotide-based, horizontal layout used by SOTA, and (ii) an integrated consensus and decoding technique that exploits the novel layout to incrementally recover data at very low sequencing coverage, and (iii) a reliable, fixed-size, block-based random access organization for DNA storage instead of a variable-sized, object-based access used by SOTA. In developing CMOSS, we make the following contributions.

\begin{itemize}
    \item Using real data from wet-lab experiments, we perform a quantification of random-access errors in DNA storage. Prior studies have done error quantification in terms of substitution, deletion, and insertion errors present in post-sequenced reads. However, there has been little focus on performing a systematic quantification of the effect of coverage bias introduced by Polymerase Chain Reaction (PCR)--the fundamental procedure used to achieve random access in DNA storage--while amplifying a complex DNA pool storing files of various sizes. We bridge this gap by presenting such an analysis (Section~
    \ref{sec:background})

    \item Using CMOSS, we perform simulation studies and two, large-scale wet-lab experiments where we encode MB-sized datasets to generate complex oligo pools. We use these experiments to (i) validate the CMOSS design by ensuring successful data recovery, (ii) compare CMOSS with SOTA in terms of read and write costs, and (iii) perform a systematic study of the impact of using PCR for randomly accessing fixed-size blocks in contrast to variable-sized objects. In doing so, we show that (i) the motif-based vertical layout and integrated consensus calling and decoding makes CMOSS resilient to errors caused by consensus bias, and (ii) the fixed-size, block-based random access organization of CMOSS makes it resilient to errors caused by PCR bias.

\end{itemize}


\section{Background}
\label{sec:background}



When used as a storage medium, DNA introduces several errors at different stages of the read and write pipelines. In this section, we will provide an overview of these errors, while making a distinction between errors that are common to all DNA storage pipelines (Section~\ref{sec:back-bias-comput}), and errors are specific to pipelines that support random access (Section~\ref{sec:back-pcr}).

\subsection{Errors due to Consensus Calling}\label{sec:back-bias-comput}

In all SOTA pipelines, binary data is stored in DNA by transforming the binary input into a quaternary sequence of nucleotides (Adenine, Guanine, Cytosine, Thymine) using an encoder. Subsequently, these sequences are utilized in the fabrication of DNA molecules, commonly referred to as oligonucleotides (or "oligos"), through a chemical process known as synthesis. The retrieval of data stored in DNA is accomplished by first sequencing the oligos to produce reads, which are quaternary sequences that correspond to the nucleotide composition of oligos. A software decoder is then used to convert the quaternary sequences into the original binary input. 

Given a set of N sequences generated by the encoder, we would ideally expect synthesis to produce N oligos, and sequencing to produce N reads, with the reads being identical to the original sequences generated during encoding. However, it is well known that both synthesis and sequencing procedures are not precise as they introduce both duplication and errors, with the extent of duplication and types of errors (substitution, insertion, and deletion) varying depending on the technology used. As a result, as the outcome of sequencing, we actually receive several noisy replicas of the original encoded sequences. 
\textcolor{black}{Thus, in all SOTA pipelines, the first step in the decoding process is to organize these sequences by clustering duplicates of the same string into distinct groups. Several clustering algorithms have been proposed for this purpose.
}
\textcolor{black}{
CD-HIT~\cite{fu2012cd} and UCLUST~\cite{edgar2010search} leverage greedy algorithms for incremental sequence clustering, offering speed at the expense of optimality. Bao et al.\cite{bao2011seed} and Antkowiak et al.\cite{antkowiak2020low}'s solutions employ advanced hashing techniques, prioritizing efficient indexing and location-sensitive clustering. Starcode~\cite{zorita2015starcode} and Meshclust~\cite{james2018meshclust} provide solutions based on precise distance calculations, with Starcode utilizing edit matrices for Levenshtein distances and Meshclust employing the mean-shift algorithm to address parameter sensitivity. The method of Jeong et al.\cite{jeong2021cooperative} focuses on Hamming distance-based clustering. Microsoft’s approach\cite{microsoft2017clustering} uses a minimal hash algorithm that enables accurate clustering over large datasets. Clover~\cite{clover2022clustering} has enhanced previous incremental algorithms in terms of accuracy and scalability, using a trie-based data structure for sequence comparison. Finally, in prior work~\cite{tldkMarinelli, onejoin}, we proposed a clustering algorithm based on embedding and locality-sensitive hashing to scale clustering over large datasets.}

After clustering, each cluster is processed separately in order to determine the most probable original sequence. This task can be formalized as follows. The original oligo $s$, of length $L$, comprises the nucleotides A, C, T, G. Given $N$ distorted copies of $s$, where each nucleotide is altered with a probability $p$ (representing deletions, insertions, or substitutions), the task is to reconstruct $s$ by finding a sequence that minimizes the total edit distance  to all given inputs. As the noisy input strings originate from the same original oligo, this challenge can be categorized as a trace reconstruction problem in information theory. \textcolor{black}{Several solutions have been proposed in literature for this problem~\cite{kannan, ViswanathanS08, Krishnamurthy, Duda, gopalan2019trace, srinivasavaradhan2021trace, sabary2023reconstruction, bar2021deep}. An example is the Bitwise Majority Alignment algorithm proposed by Batu et al.~\cite{batu-bma} which is used for oligo reconstruction in prior DNA storage experiments~\cite{organick2018random}. This algorithm effectively reconstructs the original string from its subsequences or traces by employing majority voting and appropriate shifts for each bit of the string.} 
\begin{figure}
    \centering
    \includegraphics[width=0.5\textwidth, trim={10cm 10cm 10cm 12cm},clip]{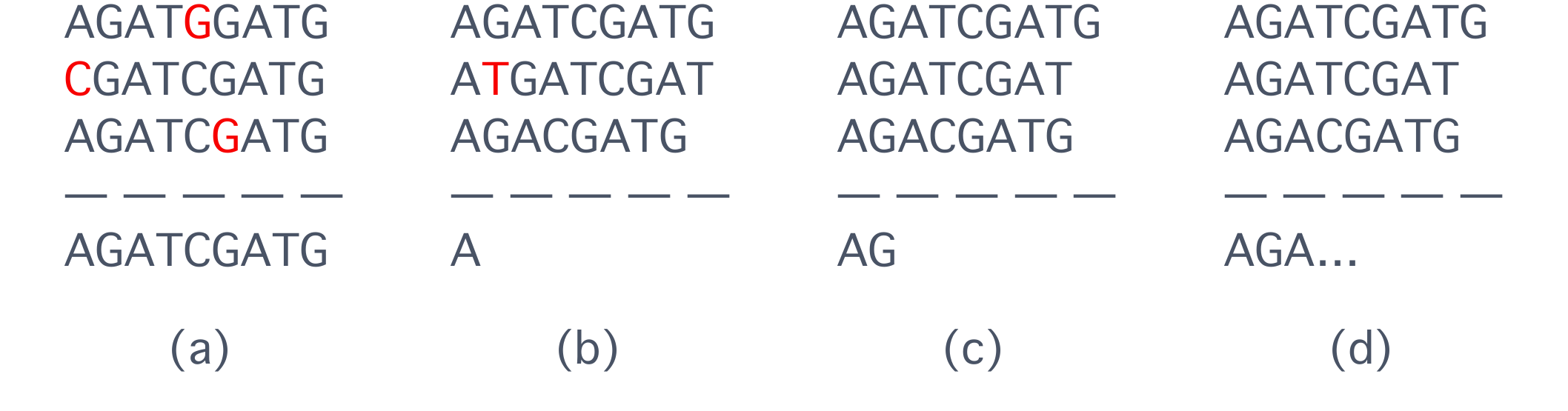}
    \caption{Example of consensus algorithm applied to a cluster of three strings in case of substitution errors only (a) and insertion/deletion errors (b)-(d).}
    \label{fig:consensus-example}
\end{figure}
Figure~\ref{fig:consensus-example} shows an example of such a reconstruction algorithm applied to a cluster of three strings. When the noisy reads contain mostly substitution errors (Figure~\ref{fig:consensus-example}(a)), and coverage (roughly defined as the number of times an oligo appears in the sequenced reads) is sufficient, we can infer the correct nucleotide at each position through majority voting. For instance, in the example, we can assume that the first nucleotide is $A$, as both the first and third strings have an $A$ as their first nucleotide. This same procedure applies to the rest of the column (nucleotides).

Handling cases with insertions or deletions is more complex. In Figure~\ref{fig:consensus-example}(c)-(d) we have the same three strings but with insertion and deletion error. When we apply consensus to the first character, as none of the strings present any error in the first position and they all have an $A$, we can assume that the first nucleotide is $A$ (Figure~\ref{fig:consensus-example}(b)). Continuing with the algorithm for the second position (Figure~\ref{fig:consensus-example}(c)), we see that the three strings differs as the first and third string contain the nucleotide $G$, while the second string has the nucleotide $T$. At this point we have to make an assumption. If we look one character ahead, we see that the second strings have the dinucleotide $GA$, similarly to the first and second strings. We can assume that the $T$ was an insertion in the second string. We can correct the insertion, by deleting the red nucleotide $T$ in the second string and and assume $G$ as nucleotide in the consensus resulting string. 

Notice that, a different consensus algorithm could have assumed something different, for example substitution error in the second string, where instead of $G$ we have $T$. Similarly, the procedure is repeated for the remaining nucleotides (Figure~\ref{fig:consensus-example}(d)). Notice that every attempt to correct the error in our strings is based on an assumption which means that there is a possibility of misinterpreting the error type. If this happens the original error and our wrong corrective attempt propagates toward the end of the reads. As the consensus calling works symmetrically whether we start the beginning or from the end of the reads, most of the errors will accumulate in the middle of the reads. Lin et al.~\cite{dnaskew-isca22} called this the reliability bias and showed that is not due to a specific consensus algorithm but it can be observed even if we use an optimal consensus algorithm. Hence, it is an intrinsic property of the trace reconstruction problem when insertion and deletion errors are present. 

%
%

The reliability bias carries significant repercussions for DNA data storage. Regarding the synthesis of DNA, as techniques improve and enable the creation of longer oligos (like Nanopore long-read sequencers), the consensus bias issue becomes more pronounced. This is because the extent of the bias is directly related to the length of the oligos. In terms of reading, while sequencing technologies are becoming more cost effective (like Nanopore long-read sequencers), they are also experiencing an increase in error rates. This trend makes the consensus algorithm less reliable due to this bias, necessitating higher sequencing coverage to effectively manage these errors. It is important to note that increased coverage translates to higher sequencing costs. 

\subsection{Errors due to Random Access} 
\label{sec:back-pcr}


Having described the reliability bias caused by consensus that affects all DNA storage pipelines, we now focus specifically on pipelines that support random access.
A single DNA pool is capable of storing several Petabytes to Exabytes of data. However, it is often necessary to retrieve only a small amount of data. Prior work has achieved this by assuming an object-based get/put interface to DNA storage and relying on the use of PCR for achieving random access of individual objects\cite{organick2018random, winston2022combinatorial}. The central idea is to associate a distinct pair of short DNA sequence, also called primers, to all oligos belonging to each distinct object. Random access is performed by using PCR to selectively amplify the DNA containing the target primer corresponding to the object that is requested. 




Prior studies have quantified the nature and frequency of substitution and indel errors introduced by different sequencing technologies and used such quantification to configure the amount of redundancy introduced during encoding/decoding\cite{heckel2019characterization}. Studies have also looked at oligo drop outs caused by coverage bias\cite{chen2020quantifying}. \textcolor{black}{Coverage bias refers to the fact that after sequencing, original oligonucleotide sequences are covered at very different rates, with some sequences being covered by multiple sequenced reads an others completely missing. Coverage bias is a well-known issue in DNA storage, with both synthesis and PCR contributing to it. During the synthesis, multiple copies of each oligo are created, with the distribution of copies being non uniform. During sequencing, a sample of synthesized DNA is extracted and PCR is used to amplify the DNA. Both samping and the inherent stochasticity of PCR can amplify the pre-existing synthesis bias leading to some oligos being copied in abundance, with others being dropped out. The issue with uneven coverage distribution lies in the inadequately represented sequences, which may not be recoverable during the decoding process because decoders typically require multiple copies of sequences to overcome randomly introduced substitution, insertion, and deletion errors. When certain sequences lack sufficient copies, it can result in the loss of information.}

However, there has been limited work on systematically quantifying coverage bias introduced by PCR-based random access\cite{winston2022combinatorial}. In practical archival scenarios, objects stored tend to have widely varying sizes. Thus, the impact of coverage bias in the context of a more realistic complex pool requires further study. In order to understand the impact of coverage bias in random access over DNA storage, we conducted a wet-lab experiment (\textbf{\labone}) where we stored three databases: SSB, TPCH, and SYN, comprising five, eight, and eight tables, respectively. The SSB and TPC-H databases were chosen from the industry-standard TPC-H benchmark\footnote{https://www.tpc.org/tpch/}, and they represent a size distribution typical in a data warehousing application. The SYN database contains randomly generated records and was configured to have table with fixed sizes. Our intention in using these databases was to isolate and study the sensitivity of PCR to the complexity of the oligo pool created by varying file sizes.

\begin{table}[htpb]
\centering
\caption{Table shows the number of oligos and the corresponding database and table primers for each table in SSB, TPCH, and SYN databases. \label{Tab:3-nb-oligos}} 
\resizebox{\columnwidth}{!}{
\color{black}{
\begin{tabular}{ccccc}
\hline
Table\# & Table Primer & \multicolumn{3}{c}{Database} \\ 
& & \multicolumn{3}{c}{(DB Primer)}\\ \cmidrule(l){3-5} 
&  &  SSB & TPCH & SYN \\
&  & (CAATG) & (GATGA) & (GTGAG) \\ \hline
1 & TTAAG & 14  & 6 &  304 \\
2 & GAATT& 16  & 18 &  312 \\
3 & AAGGT& 42  & 18 &  302 \\
4 & ACAGA& 2594  & 10 &  302 \\
5 & AGAGA& 34  & 20 &  298 \\
6 & CAGTT& & 14  & 300 \\
7 & CATAC& & 34 & 298 \\
8 & CGATA& & 16  & 306 \\	
\hline
\end{tabular}
}
}
\end{table}

\begin{figure}[htbp]
    \centering
    \includegraphics[width=0.5\textwidth]{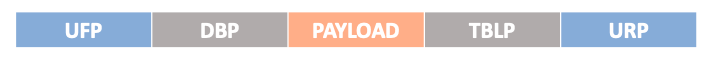}
    \caption{The oligo structure for object based abstraction. \textcolor{black}{UFP: universal forward primer, DBP: database primer, TBLP: table primer, URP: universal reverse primer. }}
    \label{fig:3-lab1-structure}
\end{figure}

The databases were converted using Goldman et al's\cite{goldman2013towards} rotational encoding approach to generate 5258 oligonucleotide sequences of 110 base pair (bp) each. The sequence consists of the payload in the middle, flanked by the database primer (DBP), table primer (TBLP), universal forward primer (UFP), and universal reverse primer (URP) on both sides (Figure~\ref{fig:3-lab1-structure}). 
Given the knowledge that the oligo length is constrained to a few hundred nucleotides due to the limitations of SOTA synthesis technology, a longer primer length translates to a reduction in payload length. To enhance the payload weight and information capacity, we utilized short primers of five nucleotides, derived from the Illumina adapter sequences. \textcolor{black}{Table~\ref{Tab:3-nb-oligos} shows the table and database primers used and the number of oligos generated for each table.}
The oligos designed in \textbf{\labone} were synthesized by Twist Bioscience and sequenced using Illumina NovaSeq. 
Due to the synthesis and sequencing errors, we observed an average error rate of 0.0033 for substitutions, 0.0003 for deletions, and 0.0003 for insertions when comparing the sequenced reads to the original oligos. 

\noindent
\textbf{Quantitative definition of coverage bias.}
We use the term ``population fraction" \cite{chen2020quantifying} to refer to the proportion of the data that belongs to a specific object (database or table) within the entire archival dataset.
Given $n$ objects, the population fraction of object $i$, denoted as $p_i$, is computed as $p_i = N_i/{\sum_{j=1}^{n}N_j}$,
where $N_i$ represents the number of sequences belonging to a specific object $i$. Given the original oligonucleotide sequences per table, we can compute the raw population fraction for each object. We refer to the raw population fraction of object $i$ as $p_i^r$. Similarly, after PCR amplification and sequencing, we can determine the population fraction of the sequenced reads. We refer to this as $p_i^s$ for object $i$. The ratio of the population fraction after sequencing to the raw population fraction is defined as \emph{population fraction change}. More formally, the population fraction change of object $i$, denoted as $c_i$, is computed as $c_i = p_i^r/p_i^s$.
Coverage is considered unbiased if and only if all sequences are equally present as the original objects' distribution. It implies that the mathematical expectation of the population fraction change for all sequences is expected to be one, indicating the absence of coverage bias.

\begin{figure}[htbp]
    \centering
    \includegraphics[width=0.35\textwidth]{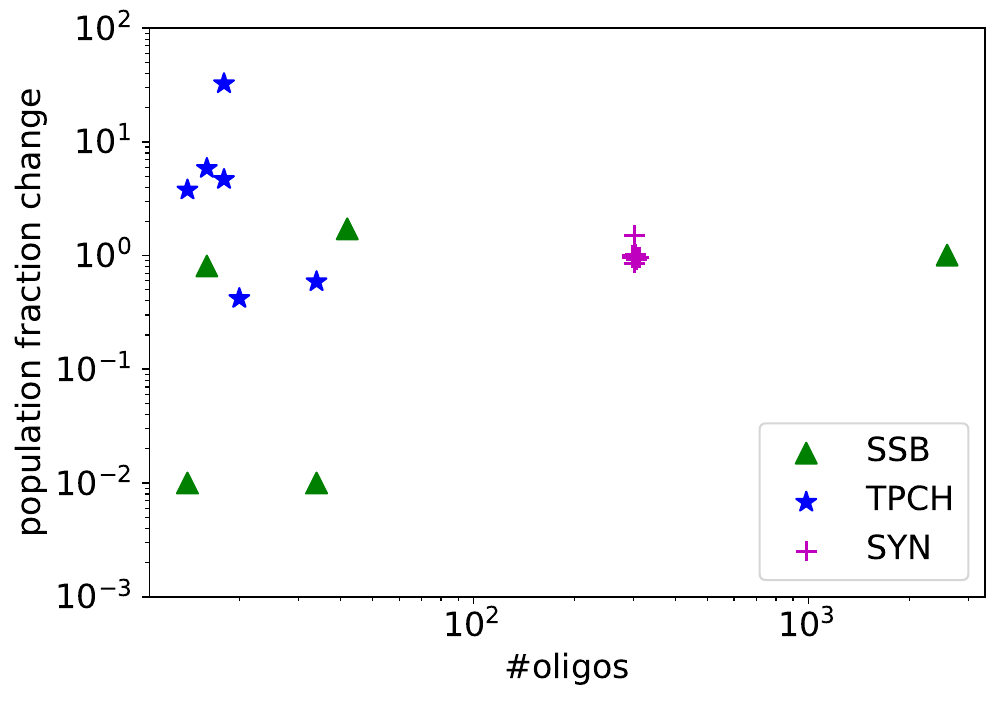}
    \caption{\textcolor{black}{The population fraction change (y-axis) and the number of oligos (x-axis) of each table in SSB, TPCH and SYN databases. The '+' purple points are overlapping together because SYN database has 8 tables of uniform size and their population fraction changes are all close to 1.}}
    \label{fig:3-lab1-fraction-change-bound}
\end{figure}

\noindent
\textbf{Coverage bias observation in real wet-lab experiment.}
To investigate whether the each database gets uniform coverage when the whole dataset is retrieved, we employed the universal forward primer to extract all the reads. Next, we aligned the reads to the oligos, using them as a reference to determine the reads' original databases with the sequence aligner tool \aname~ \cite{yan2021accel, yan2022optimizing}. The numbers of original oligos and the numbers of sequenced reads belonging to each database are presented in Table~\ref{Tab:c4-uni-uni-bias}. It illustrates that coverage is uneven and biased, as some databases are overrepresented, like SYN, while others become underrepresented, like TPCH, after sequencing. 
To study coverage bias under random access, we employed the universal forward primer in conjunction with a database primer to extract oligos belonging to one particular database. Then, we mapped the reads to oligos using {\aname} to determine their original tables. Using this alignment, we investigated the population fraction change per table. Figure~\ref{fig:3-lab1-fraction-change-bound} shows the number of oligos and the population fraction change of each table across the three databases. We can clearly see that databases with tables of varying sizes, and hence a varying number of oligos, exhibit a huge variation in population fraction change. For instance, for the SSB database, the tables vary in size from 14 oligos to 2594 oligos. The average population fraction change, which ideally should be 1, is 0.71, with smaller tables being significantly under represented (minimum population fraction change of 0.01), and some being over represented (maximum population fraction change of 1.72). The standard deviation, which ideally should be 0, is 0.73. For the TPCH database, it is even worse, with a minimum population fraction change of 0.42, maximum of 32.56 (significant over representation), average of 6.20, and standard deviation of 10.86. In contrast to these two databases, the simulated SYN database with its uniform table sizes exhibits a more uniform distribution of popular fraction change, with a minimum value of 0.86, maximum of 1.5, average of 1.03, and a standard deviation of 0.19.

\begin{table}[htpb]
\centering
\caption{The number of oligos (\#oligos), raw population fraction (raw pop frac), number of sequenced reads (\#reads), population fraction (pop frac) and fraction change (frac change) for SSB, TPCH and SYN databases.} \label{Tab:c4-uni-uni-bias} 
\resizebox{0.5\textwidth}{!}{
\begin{tabular}{rrrrrrrr}
\hline 
Database & \#oligos & raw pop frac &  \#reads & pop frac & frac change \\
\hline 
SSB & 2700 & 0.514 & 654335 & 0.388 & \textbf{0.76} \\
TPCH & 136 & 0.026 & 19576 & 0.012 & \textbf{0.45} \\
SYN & 2422 & 0.461 & 1013152 & 0.601 & \textbf{1.30} \\
\hline
\end{tabular}
}
\end{table}

\textcolor{black}{These observations expose a natural limitation of naively storing objects on DNA and using PCR-based random access without considering their sizes. Coverage bias can result in drastically different coverage for objects of different sizes, and smaller objects can get significantly under represented. Such under representation can in turn affect the effectiveness of random access by making smaller objects difficult to retieve without substantially higher redundancy to prevent data loss. In contrast, the simulated SYN databases shows that using uniformly-sized units of storage can substantially minimize this bias and ensure a more uniform representation of oligos.}
 
\begin{figure*}[ht]
    \begin{minipage}{0.5\textwidth}
    \includegraphics[width=\textwidth, trim={0 0 22cm 0}, clip]{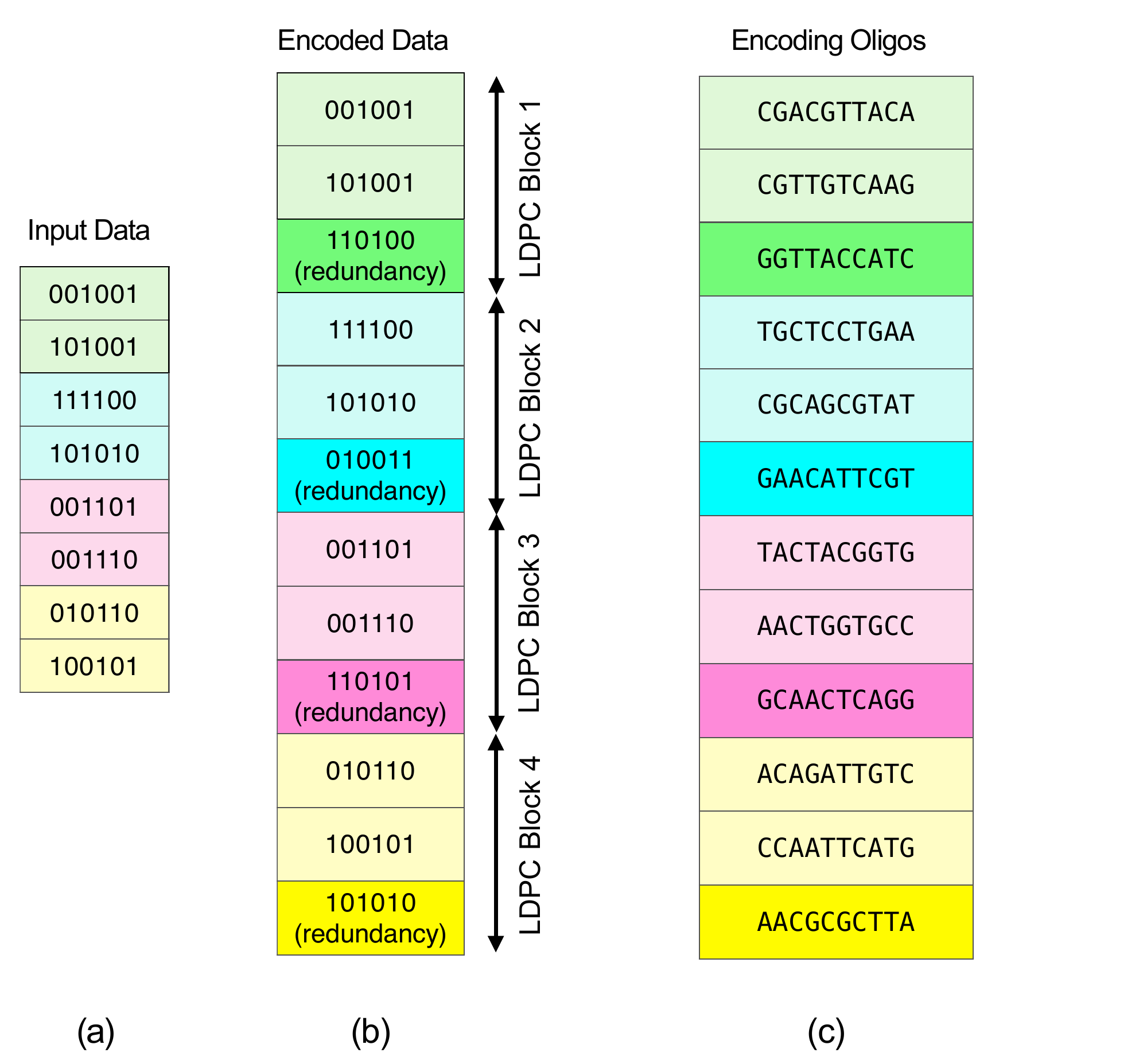}
    \end{minipage}
    \begin{minipage}{0.5\textwidth}
    \includegraphics[width=\textwidth, trim={21.5cm 0 1cm 0}, clip]{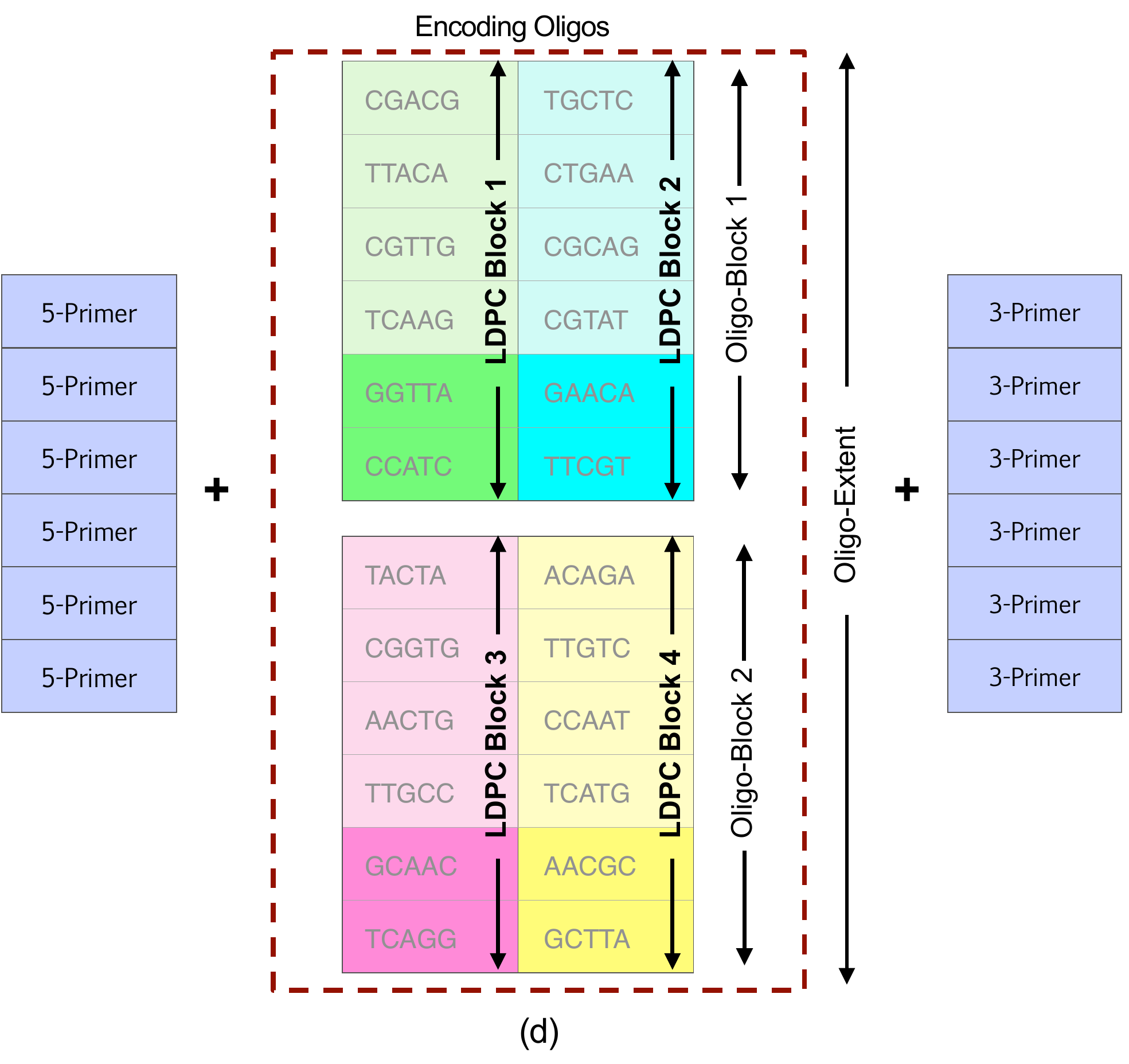}
    \end{minipage}
	\caption{ Example comparing the mapping layout of SOTA approach to our CMOSS. (a) and (b) are common to both SOTA and CMOSS. (a) The raw input data are grouped into 4 blocks as highlighted with different colors. (b) Each block in the example contains 12 bits and is encoded using LDPC error correcting code, which add 6 parity bits to each block; the resulting block is then split in 3 smaller chunks (2 containing 12 bits of data and 1 containing the parity bits).(c) SOTA encodes each chunk with one encoding oligo. As a result, each LDPC block is mapped to 3 oligos (with the same color in the picture). (d) On the contrary, CMOSS maps each block using a motif-based approach. Every group of 3 bits maps to a motif (short oligo) of 5 nucleotides. As a result, every chunk of 12 bits maps to two motifs, that are disposed vertically to form a column until the full LDPC block is encoded. Every LDPC block mapped into a column of motif is appended to the previous one: for example the blue column mapping the blue LDPC block is appended to the green column mapping the green LDPC block. Once the desired length for the oligo is reached (2 columns in the example), a new group of columns is started (in the picture, the pink and yellow columns). We refer to a column group as Oligo-Block. We call the set of Oligo-Blocks as Oligo-Extent. This organization facilitates the indexing, as every extent is identified by a pair of primers while oligos across oligo-blocks are identified with indexes. 
    Notice that for sake of simplicity the numbers reported in the figure are limited to this specific example. They are customizable, and in the actual design, we use a LDPC blocks containing 256000 bits, a motif length of 16-nts and groups of 30 bits mapping to a motif.  
    }
    \label{fig:oligo-layouts}
\end{figure*}

\section{Design}
\label{sec:design}
Having described the reliability and coverage bias issues of DNA storage, we now present CMOSS.
Our approach to archiving data in CMOSS differs from SOTA based on the key observation that the separation of consensus and decoding is a direct side-effect of the data layout, that is the way oligos are encoded. Mapping a coded block of data to a group of oligos results in that group becoming a unit of recovery. Thus, before data can be decoded, the entire group of oligos must be reassembled by consensus, albeit with errors. The key idea in our system is to change the layout from the horizontal, row-style SOTA layout (Figure~\ref{fig:oligo-layouts}(c)) to a vertical, column-style cross-oligo layout (Figure~\ref{fig:oligo-layouts}(d)). 
Our DNA storage system encodes and decodes data vertically across several oligos instead of horizontally. The key benefits, as we show later in this section, are the fact that (i) it can merge decoding and consensus into a single step, where the error-correction provided by decoding is used to improve consensus accuracy, and the improved accuracy in turn reduces the burden on decoding, thereby providing a synergistic effect, and (ii) it naturally leads to a low-coverage-bias random access organization where each unit of random access is a fixed size extent instead of a variable-sized object. In the rest of this section, we will explain the design of our system and these advantages in more detail by presenting its read and write pipelines.

\subsection{Write Pipeline}
\label{oa-dsm-write-pipeline}

\begin{figure*}[ht]
\centering
    \includegraphics[scale=0.25]{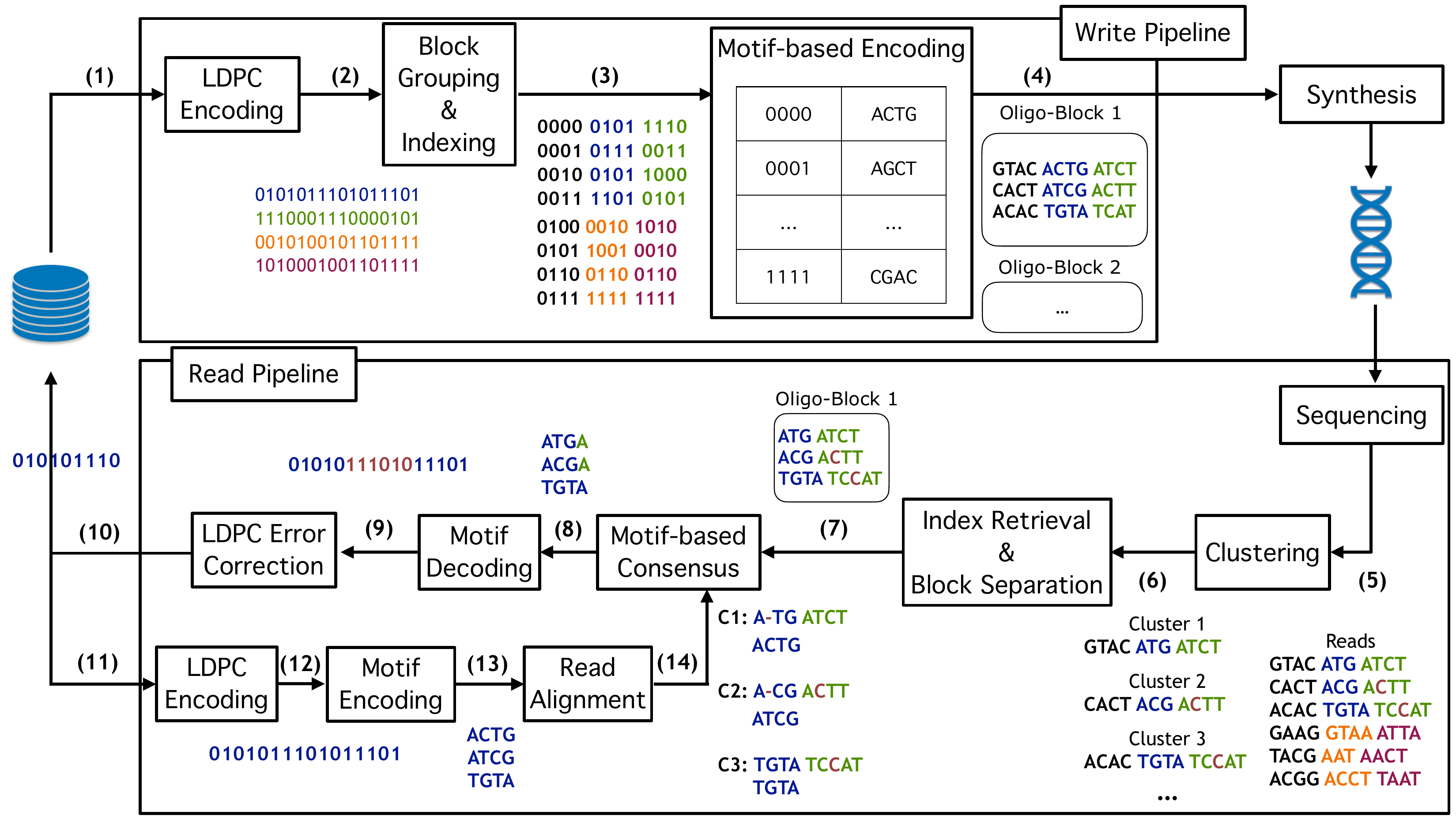}
	\caption{Our CMOSS data writing pipeline (top) shows the binary to DNA encoding pathway with a small example. (1) Binary data are split in blocks and every block is encoded using LDPC error-control code. In the example, every colored line in (2) is a LDPC-block. LDPC blocks are grouped together column-wise to form oligos blocks; they are displaced vertically in groups of 4 bits in the example. Finally, every line in the new layout is indexed (black bits in the figure) as shown in (3). Each column of LDPC block is mapped to nucleotides next: groups of 4 bits in the LDPC block are converted into motifs of 4 nucleotides passing through an associative array mapping values (of 4 bits) to motifs (4). The motifs are displaced such that every column encodes a different LDPC block. 
    Finally, the two oligos-blocks are synthesised as actual DNA strands. 
    Our CMOSS reading pipeline (bottom) shows DNA to the binary decoding pathway. For sake of simplicity, the example assumes sequencing coverage 1x. The first step in decoding is sequencing (5). Then, a clustering algorithm is applied to the noisy reads (6). As the example assumes coverage 1x, every cluster has one read only (6). Within each cluster, we apply consensus only to the first motifs (black nucleotides in the figure), in order to retrieve the indexes. Using the indexes, the clusters are further separated into their corresponding oligos-blocks (7). Then, each oligo-block enters a loop where a set of operations in each iteration is applied to each of its column of motifs. First, a motif based consensus is applied to the first column of motifs (the blue nucleotides) for every cluster. As in this example we have one noisy read per cluster, the motif-consensus will give as result for each cluster the first motif of the read itself. Because of a deletion error in the reads of cluster 1 and cluster 2, two motifs will contain the wrong nucleotide as highlighted with the green nucleotides in (8). Motifs are decoded into bits (9) and error corrected (10). The decoded bits from (10) are sent back to the LDPC encoder (11): the corrected bits are re-encoded as it was done in step (1)-(2). Then, the encoded bits are fed to the motif-encoder as it was done previously in step (4). The corrected version of the motifs in (13) are used to align reads in their corresponding clusters. Because of the realignment (14) we can stop errors to propagate in next motifs. In the example, due to the realignment we can spot two deletions that otherwise we would not be able to identify and shift the next nucleotides (the green nucleotides) accordingly. Then the whole process starting from motif consensus is repeated for the next column of motifs (green nucleotides) until all columns in one oligo-block are processed.}
    \label{fig:dsm-rd-wt-pipeline}
\end{figure*}

The top half of the Figure~\ref{fig:dsm-rd-wt-pipeline} shows the data writing pipeline of CMOSS. The input to the write pipeline is a stream of bits. Thus, any binary file can be stored using this pipeline.
The first step in processing the input involves grouping it into chunks of size 256,000 bits. Each chunk of input is then randomized. We use pseudo-randomly generated values in XOR with the bits of each block. The objective is to make the encoding oligos as far as possible from each other. This will improve the accuracy of read clustering in the data decoding stage as explained in Section~\ref{sec:oa-dsm-decoding}.
After randomization, error correction encoding is applied to each chunk to protect the data against errors. Because of the encoding, a chunk will function as unit of error control becoming the smallest recovery unit in CMOSS. The choice of code is orthogonal to the design of CMOSS and any large-block length code can be used to add redundancy. In our system, we support both Reed-Solomon (RS) and Low-Density Parity Check (LDPC) codes. We parameterize the RS code with the same block length and symbol size  as used by Organick et al.~\cite{organick2018random}(65,536 symbols with 16 bits per symbol) in their work on random access in DNA storage. We paramaterize the LDPC code with a chunk size of 256,000 bits, similar to prior work from Chandak et. al.~\cite{ldpc-dna}, which has demonstrated that such a large-block-length LDPC code is resilient to both substitution/indel errors, that cause reads to be noisy copies of original oligos, and synthesis/sequencing-bias-induced dropout errors, where entire oligos can be missing in reads due to lack of coverage~\cite{ldpc-dna}. For the rest of this section, we will only use the LDPC code to discuss the rest of the stages. In Section~\ref{sec:col-vs-row}, we present an evaluation of CMOSS with both LDPC and RS codes.

The LDPC encoded bit sequence is fed as input to the \emph{oligo-encoder} which converts bits into oligos. While SOTA approaches design each oligo as a random collection of nucleotides, our oligo-encoder designs oligos using composable building blocks called \emph{motifs}. Each motif itself is a short oligo that obeys all the biological constraints enforced by synthesis and sequencing. Multiple motifs are grouped together to form a single oligo. We use motifs rather than single nucleotides as building blocks because, as we will see later in Section~\ref{sec:oa-dsm-decoding}, integration of decoding and consensus relies on alignment which cannot be done over single nucleotide. 

In order to perform the conversion of bits into motifs, the oligo-encoder maintains an associative array with a 30-bit integer key and a 16 nucleotide-length (nt) motif value. This array is built by enumerating all possible motifs of length 16nt (AAA, AAT, AAC, AAG, AGA...) and eliminating motifs that fail to meet a given set of biological constraints. We configure our encoder to admit motifs that have up to two homopolymer repeats (AA,CC,GG, or TT), and GC content in the range 0.25 to 0.75. With these constraints, using 16nt motifs, out of $4^{16}$ possible motifs, we end up with  1,405,798,178 that are valid. By mapping each motif to an integer in the range 0 to $2^{30}-1$, we can encode 30-bits of data per motif. Thus, at the motif level, the encoding density is 1.875 bits/nt. Under the same biological constraints, we can increase this density by increasing motif size. However, we limited ourselves to this configuration due to two reasons: (i) memory limitation of our current hardware, as the current associative array itself occupies 100GB of memory, (ii) the motif design is orthogonal to the vertical encoding which is the focus of this work. 

While Figure~\ref{fig:oligo-layouts}(d) shows all columns of motifs storing only the LDPC blocks, a small subtlety in the practical implementation is that the first column in every oligo-block is dedicated to storing indexing information that orders oligos during encoding, and hence enables reordering during decoding.

The second major difference of our approach to SOTA is the layout of motifs that spread vertically in columns across a set of oligos. 
The motifs generated from an error-control coded data block are used to extend oligos by adding a new column as shown in Figure~\ref{fig:oligo-layouts}(d). This process is repeated until the oligos reach a configurable number of columns after which the process is reset to generate the next batch of oligos again from the first column. We refer to such a batch of oligos as a \emph{oligo-block (OB)}. OB is the minimum granularity of decoding in our MOSS pipeline, as all columns (i.e LDPC blocks) belonging to an OB must be encoded before starting a new OB, and similarly, all columns (i.e LDPC blocks) belonging to an OB must be decoded in order from left to right to guarantee successful data recovery as we will see in Section~\ref{sec:oa-dsm-decoding}. 

In order to scale to large datasets, we designed CMOSS as a flexible, hierarchical DNA storage system. For this purpose, we group one or more OB into a higher-level structure, termed \emph{oligo-extent (OE)}. If an OB is the unit of encoding and decoding in our MOSS, an OE is the unit of random access. Each OE is designed to be a self-contained, fixed-size (configurable during encoding) and addressable DNA storage partition. Hence, each OE is made randomly addressable by adding an unique pair of primers at two extremities of every oligo in all OBs within that OE. All oligos within the OE are addressed using the indexing strategy as described earlier. 

This hierarchical storage approach offers several advantages. First, it facilitates the storage of exceptionally large files while maintaining a relatively low primer count, as the number of primer pairs required is proportional to the number of OE, which, in turn, can be reduced by simply grouping more OB into a single OE. Second, it makes it possible to pick random access granularity during design time ranging from one OB to multiple OB. For the former, one has to set OE to be the same size as one OB. This would make the number of OE and number of OB identical. Thus, one could divide the list of primers into an equal combinatorial left--right set, and use the left primer to identify an OE and the right primer to identify an OB within each OE. For the latter, one would set the OE to span multiple OB. In this case, the left and right primers would be used to randomly access a full OE. Third, the use of OE as the granularity of random access is effectively equivalent to partitioning the DNA into fixed-sized storage units. As described in Section~\ref{sec:back-pcr}, the use of fixed-sized, extent-based random access will reduce the impact of PCR coverage bias.



\subsection{Read Pipeline}
\label{sec:oa-dsm-decoding}

Data stored in DNA is read back by sequencing the DNA to produce reads, which are noisy copies of the original oligos that can contain insertion, deletion, or substitution errors. 
Due to the hierarchical structure, decoding begins by first grouping the reads based on their OE. 
To accomplish this, the reads are aligned at both ends to unique primer pairs that designate each OE. 
Since data within different extents are independent, decoding can proceed concurrently across multiple OE, considerably speeding up the operation. For sake of simplicity, in the rest of this section we focus on decoding of a single OE with a single OB.

Recall that an OB consists of multiple oligos organized as several columns of motifs. As each oligo can be covered by multiple reads, the first step in decoding is clustering to group related reads together. 
In prior work, we developed a string clustering solution for this purpose~\cite{oa-tlkds}. Our solution uses the CGK randomized embedding algorithm~\cite{cgk-embedding} to map reads into embedded reads such that the hamming distance of embedded reads closely approximates the edit distance of the original reads. Next, our algorithm uses Locality Sensitive Hashing~\cite{lsh} to separate embedded reads into clusters based on hamming distance. Due to randomization during encoding, reads corresponding to the same original oligo are ``close'' to each other in edit distance despite errors and very ``far'' from the reads related to other oligos. Thus, the embedded hamming distances will also reflect this disparity. Our algorithm exploits this to avoid computing pair-wise edit distance which is computationally expensive. The output of this algorithm is a set of clusters, each corresponding to some unknown original oligo. While we use our solution, we would like to mention that any other read clustering solution~\cite{rashtchian2017clustering, Shinkar2022clustering} can also be used, and is orthogonal to the work presented here.

After the clustering stage, other SOTA methods apply consensus methods in each cluster to infer consensus oligos from reads. This is then followed by decoding using the consensus oligos. During decoding, SOTA methods use  error-correction codes to recover from any residual errors that might be present after consensus. Thus, decoding produces the original input bits. It is important to note that SOTA methods do not use the decoded bits from one error-control block to improve the decoding of further downstream encoded blocks. 
To explain this with an example, let us consider the first three oligos (in green) in Figure~\ref{fig:oligo-layouts}(c). Those oligos encode the LDPC-block-1 shown in Figure~\ref{fig:oligo-layouts}(b). In order to decode and correct this block of bits, SOTA approaches need to first perform consensus calling to infer the first three full oligos; then, they can convert the inferred oligos into encoded bits, and finally perform decoding with error-control codes to recover back the original input bits. However, once the original bits for the green block are retrieved, they will be only used as part of the final output, that is the reconstructed original input file.

In CMOSS, we exploit the motif design and columnar layout of oligos to progressively perform consensus and decoding in an integrated fashion as shown in the bottom part of Figure~\ref{fig:dsm-rd-wt-pipeline}. Unlike other approaches, our system processes the reads one column at a time. Thus, the first step is motif-based consensus which takes as input the set of reads and produces the first column of motifs. The choice of consensus algorithm is orthogonal to CMOSS design. We use an alignment-based algorithm that we developed previously for motif-based consensus calling~\cite{oa-tlkds}. The algorithm works by taking a motif-length portion of each read belonging to a cluster, aligning them, and taking a position-wise majority of aligned motifs to determine a consensus motif. As each cluster corresponds to an oligo, this process is done for each cluster to determine one consensus motif per cluster, and hence all consensus motifs for the first column of motifs. These motifs are then fed to our \emph{CMOSS oligo-decoder}, which is the inverse of the encoder, as it maps the motifs into their 30-bit values. 
Note here that despite consensus, the inferred motifs can still have errors. These wrong motifs will result in wrong 30-bit values. These errors are fixed by the \emph{LDPC-decoder}, which takes as input the 30-bit values corresponding to one LDPC block and produces as output the error-corrected, randomized input bits. These input bits are then derandomized to produce the original input bits for that block.
Different from SOTA, in CMOSS, these decoded bits are reencoded again by passing them through the LDPC-encoder and the oligo-encoder. This once again produces the correct first column of motifs as it would have been done during encoding in the write pipeline. The correct motifs are then used to realign reads within each cluster so that the next round of decoding for the second column start at the correct offset. This whole process is repeated for all subsequent columns.

\begin{figure*}[t]
\centering
    \includegraphics[height=3.5cm, trim={0 0 0 0}]{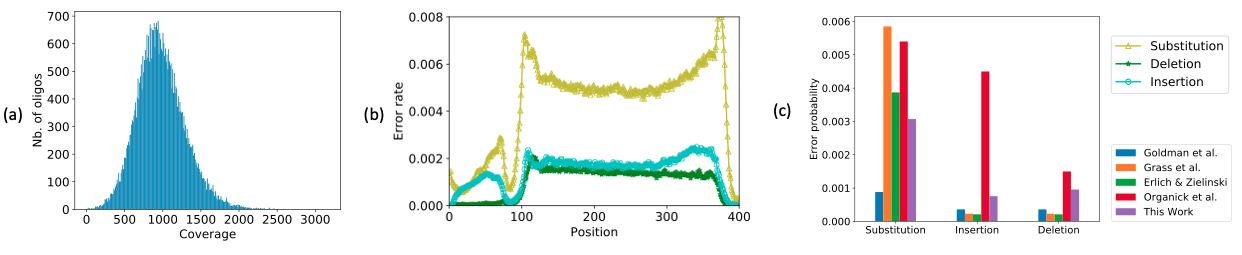}
	\caption{Statistics about \textbf{\labtwo} where the input was encoded by CMOSS with 30\% LDPC redundancy, the oligos were synthesized by Twist Biosciences and sequenced by the Oxford Nanopore PromethION. (a) The histogram of coverage across oligos. The x-axis is the sequence coverage, y-axis is the number of oligos having that sequence coverage. (b) The substitution, deletion and insertion error rates of each position of the read. (c) Comparison of the overall average substitution, deletion and insertion errors rates with previous work. 
 }
    \label{fig:exp2-all}
\end{figure*}


The intuition behind this realignment is as follows. An insertion or deletion error in the consensus motifs will not only affect that motif, but also all downstream motifs also due to a variation in length. For instance, if we look at the example sequenced reads in step (6) of Figure~\ref{fig:dsm-rd-wt-pipeline}, we see a deletion error in the first read $GTACA-TGATCT$ which should have been $GTACA\textbf{C}TGATCT$ (according to the oligos generated in step (4) of the same figure). This results in the second motif being incorrectly interpreted as $ATGA$ instead of $ACTG$. Left uncorrected, this error will spill over to the third motif which will be read as $TCT...$ (instead of $ATCT$). Thus, with SOTA approaches, an error early in a consensus oligo keeps propagating leading to the reliability bias as explained in Section~\ref{sec:back-bias-comput}. On the contrary, in CMOSS, every column stores a full LDPC block, and we decode one column at a time. Thus, we can use the decoded bits from one column to regenerate the correct motifs by rencoding them during decoding.  We can use the correct motifs to fix these errors by realigning the correct motifs against reads. This realignment will determine the position where current motif ends and the next motif begins, and hence, determine the starting point for the next column. As a result, any consensus errors in one column can be fixed by realignment and do not propagate downstream limiting the impact of positional bias. 

Note here that this realignment is only possible because we use motifs, as two sequences can be aligned accurately only if they are long enough to identify similar subsequences. Thus, columnar layout without motifs, or with just nucleotides, would not make realignment possible. Similarly, integrating consensus and decoding is possible only because of the columnar layout, as the SOTA layout that spreads a LDPC block across several oligos cannot provide incremental reconstruction.

Finally, a small refinement to the decoding procedure we have described so far is the special handling required to deal with indexing information for the case when an OE has more than one OB. Recall that OE is the unit of random access, while OB is the unit of decoding. Also recall that we store indexing information in the first column of motifs across the whole OE. Thus, in the case where an OE has more than OB, primers are used to identify OE, while this index is used to indirectly identify the OB of each oligo within an OE. So the decoding of the first column of motifs is special in that it produces this indexing information across all OBs within an OE. The index information is used to separate reads into constitutent OBs and starting from the second column, we switch to per-OB processing. This whole process is illustrated with a simple example in Figure~\ref{fig:dsm-rd-wt-pipeline}.



\section{Evaluation}
\label{sec:evaluation}




In this section we present a thorough evaluation of CMOSS. First, we present the results from two wet-lab experiments to study the ability to achieve full data recovery (Sec.~\ref{sec:wetlab-val-exp2} and Sec.~\ref{sec:wetlab-val-exp3}). Next, we compare CMOSS with various SOTA approaches with respect to read cost and write cost to show that our design can lead to substantial cost reduction (Sec.~\ref{sec:oadsm-vs-sota}). Following this, we isolate and analyze the advantage of using a column-based design by comparing CMOSS with a row-based pipeline (Sec.~\ref{sec:col-vs-row}). Finally, we present a large scale simulation to validate our end-to-end CMOSS pipeline on a scaled up dataset (Sec.~\ref{sec:larg-scale}). 
We conduct all the experiments on a local server equipped with a 12-core CPU Intel(R) Core(TM) i9-10920X clocked at 3.50GHz, 128GB of RAM. The core components shown in Figure~\ref{fig:dsm-rd-wt-pipeline} have been implemented in C++17. 


\subsection{Small-Scale Wet-Lab Validation}
\label{sec:wetlab-val-exp2}

As the first prototype test, we used the TPC-H DBGEN utility to generate a compressed database of 1.2MB which was subsequently encoded by CMOSS, configured with 30\% LDPC redundancy, into 44376 oligos of length 200nt partitioned into sixteen OEs. The oligos were synthesized by Twist Biosciences. 
Subsequently, we sequenced the oligos using the Oxford Nanopore PromethION platform with Ligation Sequencing Kit V14 (SQK-LSK114), producing a total of 43M reads (\textbf{\labtwo}). 


\subsubsection{Error Pattern}

\begin{figure*}[t]
\centering
    \includegraphics[height=3.5cm, trim={0 0 0 0}]{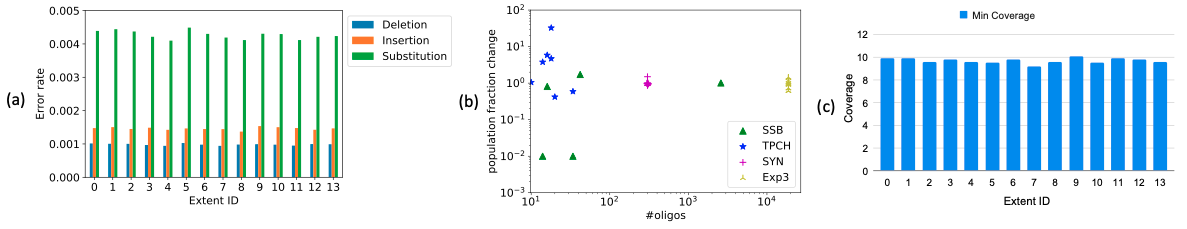}
	\caption{Statistics about \textbf{\labthree} where the input was encoded by CMOSS with 10\% LDPC redundancy, the oligos were synthesized by Twist Biosciences and sequenced by the Oxford Nanopore PromethION. 
 (a) The error rate of the reads selected by each extent's primers. 
 (b) The population fraction change (y-axis) and the number of oligos (x-axis) of each extent. The yellow points are overlapping together because \textbf{\labthree} has 14 extents of uniform size and their population fraction changes are all close to 1. We also plotted those metrics about \textbf{\labone} mentioned in Figure~\ref{fig:3-lab1-fraction-change-bound} for comparison. 
 (c) The minimum coverage required for full data reconstruction per extent.}
    \label{fig:exp3-all}
\end{figure*}

To perform error characterization, we aligned the sequenced reads generated from \textbf{\labtwo} to the original oligos using Accel-Align~\cite{yan2021accel} sequence aligner. 99.9999\% reads were aligned to a reference oligo, indicating a very high quality of the generated read set. Figure~\ref{fig:exp2-all}.a shows the coverage histogram and it can be observed that each reference oligo is covered by at least one read, with a median coverage of 951$\times$, minimum coverage of 5$\times$, and a maximum coverage of 2500$\times$. We deliberately sequenced the oligos at such high coverage to test recovery at various coverage levels as we present later. 

The average error rates are 0.003 for substitutions, 0.0008 for deletions, and 0.001 for insertions computed by BBmap \cite{bushnell2014bbmap}. The error rate per position is illustrated in Figure~\ref{fig:exp2-all}.b. Note that while the data-carrying payload had a length of 160, our reads are longer as they include the primers that were appended at both ends of the oligo for sequencing. 
As these primers get trimmed out during read preprocessing, the error rate of relevance to us is the middle portion of the read which corresponds to the encoded, data-carrying portion of the oligo. We see that in this portion, the substitution rate is dominant, which is 3$\times$ higher than insertion and deletion rates. Figure~\ref{fig:exp2-all}.c compares our error rates with those reported in prior work on DNA storage \cite{goldman2013towards, grass2015robust, erlich2017dna, organick2018random, heckel2019characterization}. We calculated these statistics using raw reads without any quality-based filtering. As can be seen, our reported error rates are lower than those reported in literature. Both array synthesis and Nanopore sequencing have improved in accuracy over the past few years, with Nanopore PromethION platform offering single-read accuracy of over 99\% with LSK114 kit\cite{nanoporetechLSK114}. It is hard to attribute a precise fraction of improvement in error rate to synthesis and sequencing without an isolated comparison of each with other studies. However, we can see that the overall trends of relative errors are similar.

\subsubsection{Data Recovery}
In order to test end-to-end decoding, we first used the full 43M read dataset generate from \textbf{\labtwo} as input to the decoding pipline. We were able to achieve full data reconstruction, given the ability of CMOSS to handle much lower coverage levels and higher error rates.
In order to stress test our decoding pipeline and identify the minimum coverage that allows fully reconstruction of data, we repeated the decoding experiment on smaller readsets which were derived by randomly sampling a fraction of reads from the 43M read dataset. In doing so, we found that CMOSS was able to perform full recovery using just 200K reads, which corresponds to a coverage of 4$\times$. At this coverage, nearly 3500 out of 44376 reference oligos were completely missing. However, the LDPC code and column-based decoding were able to successfully recover data. As further reduction in coverage led to data loss, we validate 4$\times$ as the minimum coverage CMOSS can handle with our wetlab experiment. Computing the costs for minimum coverage, we get a read cost of 3.37 nts/bit, and a write cost of 0.72 nts/bit.

\subsection{Large-Scale Wet-lab Validation}
\label{sec:wetlab-val-exp3}

As a large-scale test of random access, we stored a 13MB tar archive containing culturally significant documents, including images, PDF files, and text documents, sourced from a national archive (\textbf{\labthree}). Employing a methodology similar to that of \textbf{\labtwo}, we encoded the input by with just 10\% LDPC redundancy, lower than the one adopted in \textbf{\labtwo}. The reason of a lower coverage is that in \textbf{\labtwo} we already proved a full reconstruction with a very low coverage at 30\% redundancy. Thus, for this experiment we tested our system with a lower redundancy overhead. 
This resulted in a total of 262,836 sequences of 240nt stored in 14 OB. For the purpose of this experiment, given the limited number of extents, we made the number of OE and OB identical. This means that with each OB and hence, each OE, store 468KB of information (15 256000 bit LDPC blocks per OB). This becomes the unit of random access. To identify each OE/OB individually, we added a 20nt 5’-primer and a 20nt 3’-primer to each sequence (for a total of 280nt oer oligo) which were synthesized with Twist Biosciences. To evaluate data recovery per extent, we conducted 14 independent wetlab experiments. Each wetlab used one extent's distinct left and right primers during PCR amplification to randomly select that extent. Subsequently, the amplified oligos were sequenced using the same Oxford Nanopore PromethION platform to produce 6.1M reads.

\subsubsection{Error Pattern}
Figure~\ref{fig:exp3-all}.a shows the average substitution, deletion, and insertion error rates of reads per extent. As can be seen, the rates are similar across extents and comparable to the results of \textbf{\labtwo} shown in Figure~\ref{fig:exp2-all}.b. 


\subsubsection{Coverage Bias}
As we explained in Section~\ref{sec:back-pcr}, 
file-based random access suffered from a high coverage bias when files are of varying sizes. To investigate bias under block-based random access with CMOSS, we aligned all 6.1M sequenced reads from \textbf{\labthree} to their original oligos using \aname~ in order to determine their original extents. We used this alignment to calculate population fraction change. Figure~\ref{fig:exp3-all}.b is an extension of Figure~\ref{fig:3-lab1-fraction-change-bound} with the points for each of the fourteen extents from \textbf{\labthree} added. As each extent has the same number of oligos, all points cluster together on the x-axis. Due to the uniform extent size, the population fraction change across all extents is close to 1, with a standard deviation of 0.278. This result is in clear contrast to TPCH and SSB database results, where population fraction change varies a lot with standard deviations of 0.7 and 10.8. The low standard deviation in \textbf{\labthree} case with CMOSS signifies 
that the oligos now have a more uniform coverage across extents after the PCR process due to the fact each unit of random access has an identical number of oligos, just like the simulated SYN database with uniform table sizes.

\subsubsection{Data Recover per Extent}
We utilized all available reads for each extent to independently reconstruct the data blocks stored within them using the CMOSS read pipeline. The average coverage of each extent is 30x, with a minimum coverage of 17x and a maximum coverage of 42x. 
We compared the decoded bits with corresponding segments of the original binary file to confirm that every segment of the file was accurately reconstructed.


To evaluate the robustness of our system, we conducted an experiment to determine the minimal read coverage required at the extent level for complete data recovery. This was achieved by progressively reducing the number of reads sampled from available pool of reads of each extent until the lowest count necessary for full extent recovery was identified. Utilizing this dataset, we calculated the minimum coverage per extent.
The findings, illustrated in Figure~\ref{fig:exp3-all}.c, reveal the minimum coverage necessary to achieve full recovery for each extent. From our analysis, we can see that the minimum coverage across all extents is around 9.5x, and this is consistent for each individual extent, demonstrating the uniformity and reliability of CMOSS. 


\subsection{SOTA Comparison}
\label{sec:oadsm-vs-sota}

\begin{table*}[htbp]

\centering
\caption{Comparison of this work with SOTA DNA Storage Methods. The table summarize the information about the wet lab experiments when are available in their publication, such as size of file encoded, number of oligos generated, oligo length, minimum coverage they reported in their publications to fully reconstruct the encoded data. From this information, we computed two metrics for comparisons: write cost, defined as $\frac{\#nts-in-oligos}{\#bits}$ and read cost $\frac{\#nts-in-reads}{\#bits}$.For both the metrics, lower is better.}

\begin{tabular}{|c|c|c|c|c|p{1.4cm}|p{1.3cm}|}
\hline
\textbf{Reference} & \textbf{Binary Size} & \textbf{Nb of Oligos} & \textbf{Oligo Length (nt)} & \textbf{Recovery Coverage}  & \textbf{Write Cost (nt/bit)} & \textbf{Read Cost (nt/bit)} \\
\hline
Church et al.~\cite{church2012next} (2012) & 658 KB & 54,898 & 115 & 3000 &  1.17 & 3513.66 \\
\hline
Goldman et al.~\cite{goldman2013towards} (2013) & 650 KB & 153,335 & 117 & 51 & 3.37 & 171.83 \\
\hline
Grass et al.~\cite{grass2015robust} (2015) & 85 KB & 4,991 & 117 & 372  & 0.84 & 311.97 \\
\hline
Bornholt et al.~\cite{bornholt2016dna} (2016) & 150 KB & 16,994 & 80 & 40 & 1.11 & 44.26 \\
\hline
Yazdi et al.~\cite{yazdi-dna-storage} (2017) & 3.55 KB & 17 & 1000 & 200 & 0.58 & 116.91 \\
\hline
Erlich and Zielinski~\cite{erlich2017dna} (2017) & 2.11 MB & 72,000 & 152 & 10.4 & 0.62 & 6.43 \\
\hline
Organick et al.~\cite{organick2018random} (2018) & 200 MB & 13,448,372 & 110/114 & 5 & 0.91 & 4.57 \\
\hline
Anavy et al.~\cite{anavy2019} (2019) & 22.5 B & 1 & 42 & 100 & 0.23 & 23.33 \\
\hline
Choi et al.~\cite{choi2019high} (2019) & 135.4 KB & 4,503 & 111 & 250 & 0.45 & 112.66 \\
\hline
 S. Chandak et al.~\cite{chandak2019improved} (2019) & 192 KB & 11,892 & n.a. & 5 & 0.78 & 4.46 \\
\hline
THIS WORK (Exp2) & 1.2 MB & 44376 & 200 & 4 & 0.70 & 2.82 \\
\hline 
THIS WORK (Exp3) & 13 MB & 262,836 &	280 & 9.5 & 0.57 & 5.49 \\
\hline
\end{tabular}

\label{tab:SOTA-comparison}
\end{table*}

\begin{figure*}[t]
\minipage{0.32\textwidth}
	\includegraphics[width=\linewidth, trim={0 1.80cm 0 1cm},clip]{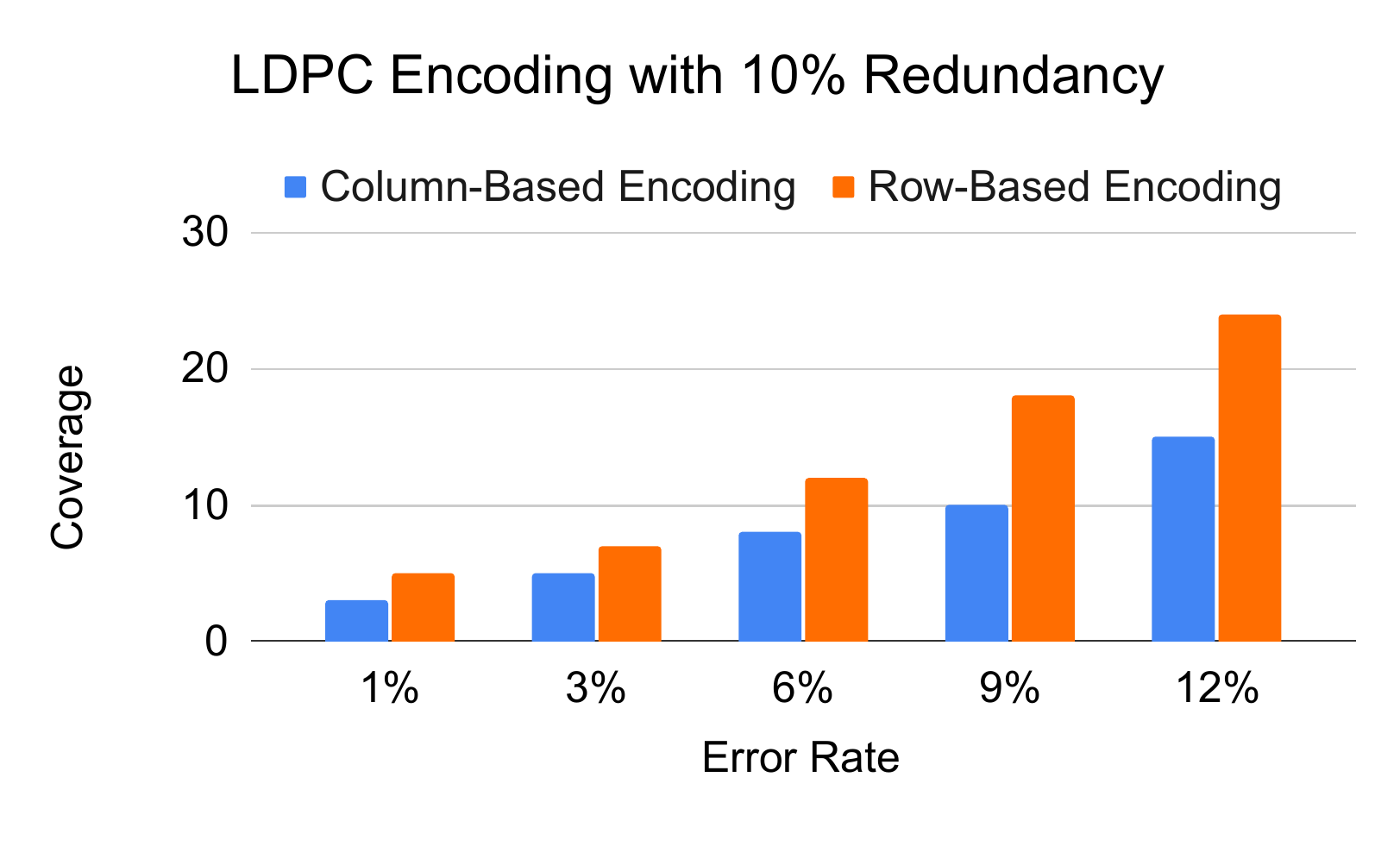}
    \caption{Minimum coverage required by our column-based and row-based implementations for decoding a 3MB archive file encoded with LDPC code at 10\% redundancy. Lower is better.\\}
    \label{fig:10p-coverage}
\endminipage\hfill
\minipage{0.32\textwidth}
	\includegraphics[width=\linewidth, trim={0 1.80cm 0 1cm},clip]{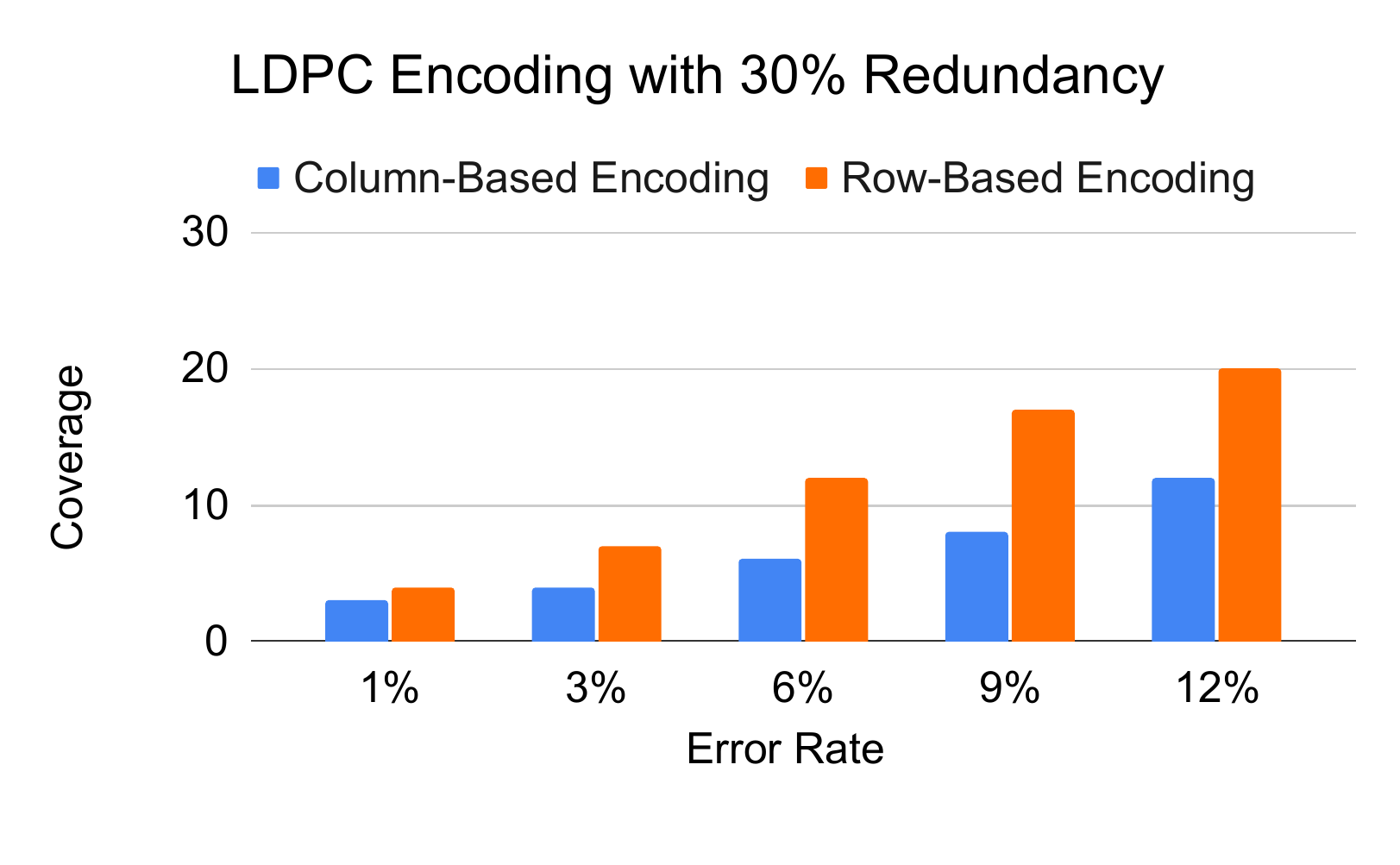}
	\caption{Minimum coverage required by our column-based and row-based implementations for decoding a 3MB archive file encoded with LDPC code at 30\% redundancy. Lower is better.\\}
	\label{fig:30p-coverage}
\endminipage\hfill
\minipage{0.32\textwidth}
	\includegraphics[width=\linewidth, trim={0 1.8cm 0 1cm},clip]{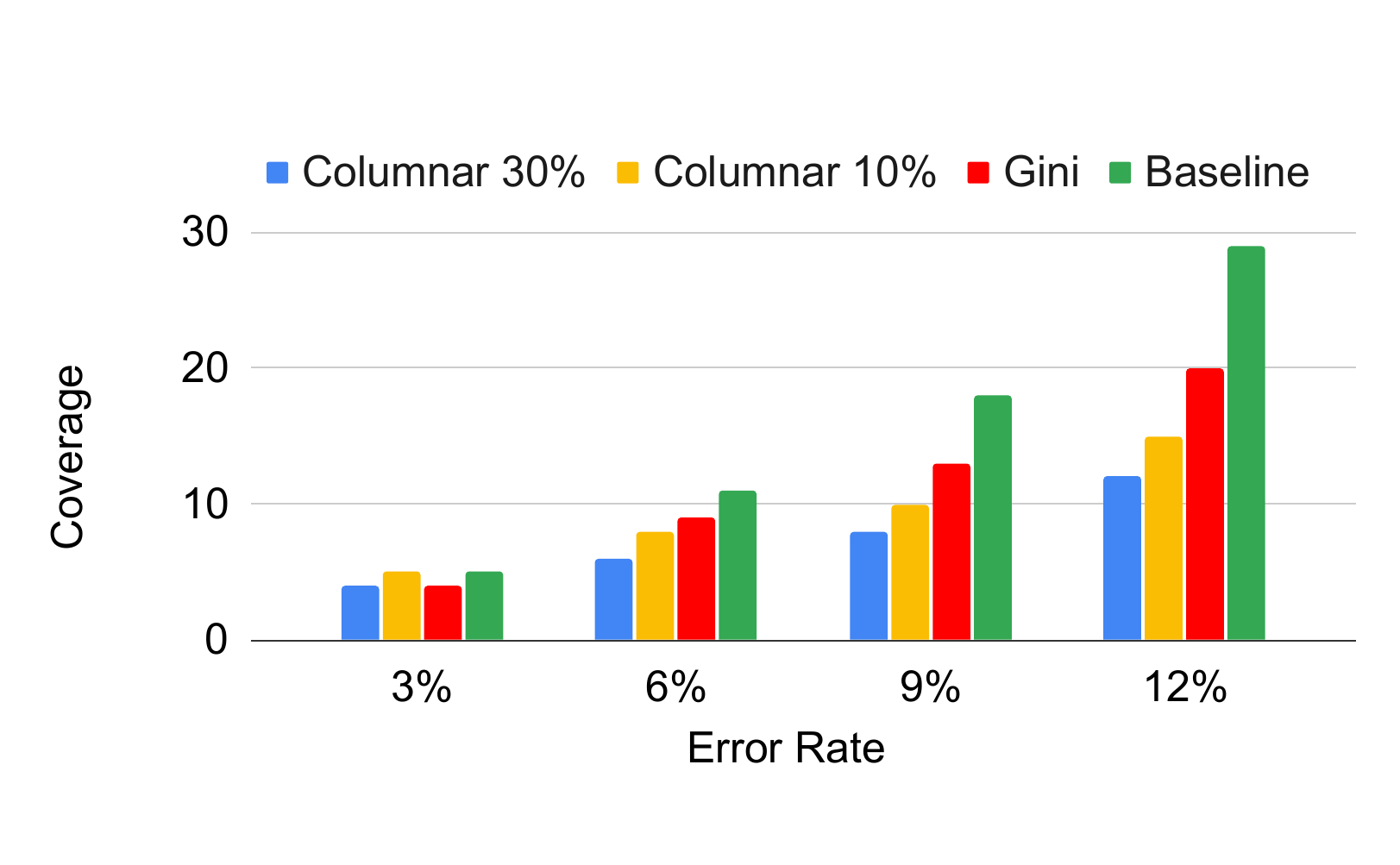}
    \caption{Comparison of this work (columnar 10\% and 30\%) with Gini and a baseline without Gini reported by Lin et al.~\cite{dnaskew-isca22}. The y-axis reports the minimum coverage required to fully decode data at various error rates. Lower is better}
    \label{fig:gini-comparison}
\endminipage\hfill
\end{figure*}

Having discussed the results from our real-world wetlab experiments, we will now present a comparison of CMOSS with SOTA approaches in terms of reading and writing cost~\cite{organick2018random,ldpc-dna,dnaskew-isca22}. Writing cost is defined as $\frac{\#nts-in-oligos}{\#bits}$, where the numerator is the product of the number of oligos and the oligo length, and the denominator is the input data size. Thus, higher the redundancy and encoding overhead, higher the write cost. The reading cost is defined by $\frac{\#nts-in-reads}{\#bits}$. The numerator is the sum total of all read lengths, and denominator is the input size. Thus, higher the coverage required, higher the read cost.

Table~\ref{tab:SOTA-comparison} shows the read and write cost for CMOSS and other SOTA algorithms. We would like to emphasize here that our goal in reporting these results is not to directly compare our work with SOTA based on these metrics; an apples-to-apples comparison is not possible given differences in all stages of the DNA storage pipeline. Rather, our goal is to position our results in the broader context. For CMOSS, we compute these costs based on \textbf{\labtwo} and \textbf{\labthree}. We only include these two results as they are from real wet-lab experiments and not simulation studies. For \textbf{\labtwo}, we compute the write cost using the 44376 oligos synthesized to encode a 1.2MB archive and for read cost the minimum number of reads (corresponding to a coverage 4x) needed to fully reconstruct the original data. Similarly, for \textbf{\labthree} we computed the write cost by considering the 262,836 oligos used to encode the 13MB archive while the read cost was based by considering the minimum coverage that allows us to fully recover the entire archive. We do not report data for \textbf{Exp.1} in Table~\ref{tab:SOTA-comparison}, as it was used to demonstrate coverage bias and did not use CMOSS to encode data. For SOTA approaches, we reproduce the costs from their publications where available. There are several observations to be made. 

First, let us compare CMOSS with row-based SOTA approach that also uses LDPC (by S. Chandak et al.~\cite{ldpc-dna}). Both these cases use the same LDPC encoder configured with 30\% redundancy. The cost reported here is for around 1\% error rate in both cases. Clearly, the CMOSS approach has both a lower write and read cost. The difference in write cost can be explained due to the fact that in the row-based LDPC approach, the authors also added additional redundancy in each oligo in the form of markers which they used in their decoder. CMOSS is able to achieve 100\% data reconstruction using the same LDPC encoder at a much lower coverage level without such markers as demonstrated by the lower read cost. 

Among SOTA, two other pieces of related work that have competitive read/write cost are the large-block RS coding by Organick et al.~\cite{organick2018random} and fountain codes by Erlich et al.~\cite{erlich2017dna}. Comparing CMOSS with these, we see that CMOSS \textbf{\labtwo} with 30\% redundancy provides better read cost than both, but worse write cost than the fountain coding approach. CMOSS \textbf{\labthree} has worse read cost than Organick et. al. but a better write cost than both as it uses 10\% redundancy. As we mentioned earlier, we can further improve the write cost for CMOSS using several approaches. First, the CMOSS results from \textbf{\labtwo} in Table~\ref{tab:SOTA-comparison} were obtained with a 30\% redundancy based on its ability to handle even 12\% error rate. For lower error rates (less than 1\%), as was the case with the Fountain coding work, even 10\% redundancy would be able to fully restore data at extremely low coverage (3$\times$ as shown in Figure~\ref{fig:10p-coverage}). Second, as mentioned in Section~\ref{sec:design}, scaling the motif set by using longer motifs (17nt and 33 bits) could allow us to increase bit-level density further from 1.87 bits/nt to over 1.9 bits/nt. These two changes would lead to further reduction in write cost without any adverse effect on the read cost. As this work was predominantly about reducing the read cost, we leave these optimizations to future work.

Lin et al.~\cite{dnaskew-isca22} recently presented the Gini architecture which interleaves nucleotides across oligos in order to minimize the impact of consensus errors. We also tried to compare CMOSS with Gini, but we could not derive the read/write cost for Gini, which was also not reported, due to lack of statistics about reads. However, as our evaluation methodology is identical to Gini, we present a direct comparison of results in terms of minimum coverage required by both approaches. Figure~\ref{fig:gini-comparison} shows the minimum coverage required by CMOSS, Gini, and a baseline without Gini reported by Lin et al.~\cite{dnaskew-isca22}, to perfectly recover data at various error rates. At 18.4\% redundancy based on RS coding, the reported baseline needed 30$\times$ to recover data at 12\% error rate. Gini, in contrast, provided a 33\% improvement as it needed a minimum coverage of 20$\times$ at 12\% error rate to guarantee full recovery. CMOSS configured at 30\% redundancy with LDPC encoding provides a 40\% improvement over Gini, it requires only 12$\times$ coverage. Comparing Figure~\ref{fig:gini-comparison} with Figure~\ref{fig:10p-coverage}, we see that CMOSS provides 25\% less coverage (15$\times$) even at 10\% redundancy compared to Gini. Thus, CMOSS has a much lower read cost, thanks to the integrated consensus and decoding enabled by column-based organization.

Finally, we would like to mention that there are other SOTA approaches that tried to add to Table~\ref{tab:SOTA-comparison}~\cite{press2020hedges, antkowiak2020low, bar2021deep, tomek2019driving, tomek2021promisc, lin2020dynamic}. But we could not find all the information necessary for computing the read and write costs. Hence, we did not report these methods in Table~\ref{tab:SOTA-comparison}.

\subsection{Benefits of Column-Based Design}
\label{sec:col-vs-row}

In order to ensure that the benefits of CMOSS are due to the column-based design and not other parameters, we have developed a row-based version of the pipeline shown in Figure~\ref{fig:dsm-rd-wt-pipeline}, where we fixed all other parameters (clustering and consensus algorithms, LDPC block size, motif set, etcetera), and only changed two aspects to make it similar to SOTA: (i) replace CMOSS encoder with row-based encoder that maps one LDPC block to multiple oligos, (ii) perform consensus to infer entire oligos first, and then decode separately. 

In order to compare the column-based and row-based pipelines, we perform an end-to-end DNA storage simulation study using both pipelines. First, we use both pipelines to generate the oligos for a 3MB TPC-H archive file (3MB size was chosen based on calculations that ensure that both pipelines produce the same number of oligos). We configure LDPC encoder to generate two datasets, with 10\% and 30\% redundancy. Then, we encode the two datasets using both pipelines, while fixing the oligo length to 50 motifs per oligo (800nt), generating four oligos datasets, two containing $18773$ oligos (row-based/column-based at 10\% redundancy), and the other two containing $22187$ (row-based/column-based at 30\% redundancy) oligos.

We compare the row-based and column-based pipelines by evaluating the minimum coverage required at 10\% and 30\% redundancy levels to achieve 100\% error-free reconstruction of the input data at various the error rates (1\% to 12\%). We conduct the experiment similar to SOTA~\cite{ldpc-dna,dnaskew-isca22} as follows. For each error rate, and for each of the four oligo sets, we generate read datasets at various coverage levels (1$\times$ to 25$\times$). In order to generate reads, we first duplicate each oligo a certain number of times according to the configured coverage level. Then we inject random errors at random positions in each read. We inject insertion, deletion and substitution with an equal probability, and the number of errors injected per read follows a normal distribution with mean set to the configured error rate. We then decode the read datasets using both pipelines and identify the minimum coverage level required to fully recover the original data. We repeated each experiments three times and the minimum coverage level remained constant for every run. We reported these values in Figure~\ref{fig:10p-coverage} and Figure~\ref{fig:30p-coverage}.



Figure~\ref{fig:10p-coverage} shows the minimum coverage for data encoded with 10\% redundancy. Clearly, column-based encoding outperforms the row-based one, as it reduces the coverage required up to 40\% for high error rates. This reduction in minimum coverage can be intuitively explained as follows. Row-based encoding maps an LDPC block into multiple oligos. This implies that a single erroneous oligo can lead to a data loss of up to 1500 bits (50 motifs per oligo $\times$ 30 bits per motif). As explained in Section~\ref{sec:design}, all that is required for an oligo loss is a single insertion/deletion error in the first motif after consensus. On the other hand, 
an oligo loss in CMOSS only causes a loss of 30 bits in each of the LDPC blocks, thanks to the column-based encoding. Further, the integrated consensus and decoding can fix consensus errors in early rounds so that they do not affect future rounds. Due to these reasons, the LDPC decoder works much more effectively when paired with column-based layout rather than row-based encoding. 
The results are similar for data encoded with 30\% redundancy as well, as shown in Figure~\ref{fig:30p-coverage}. Notice that in the 30\% case, both row-based and CMOSS pipelines have a minimum coverage lower than the 10\% redundancy case. This is expected, as a higher redundancy implies a higher tolerance to errors. 

\begin{figure}[h]
    \minipage{0.45\textwidth}
	\includegraphics[width=\linewidth, trim={0 1.8cm 0 1cm},clip]{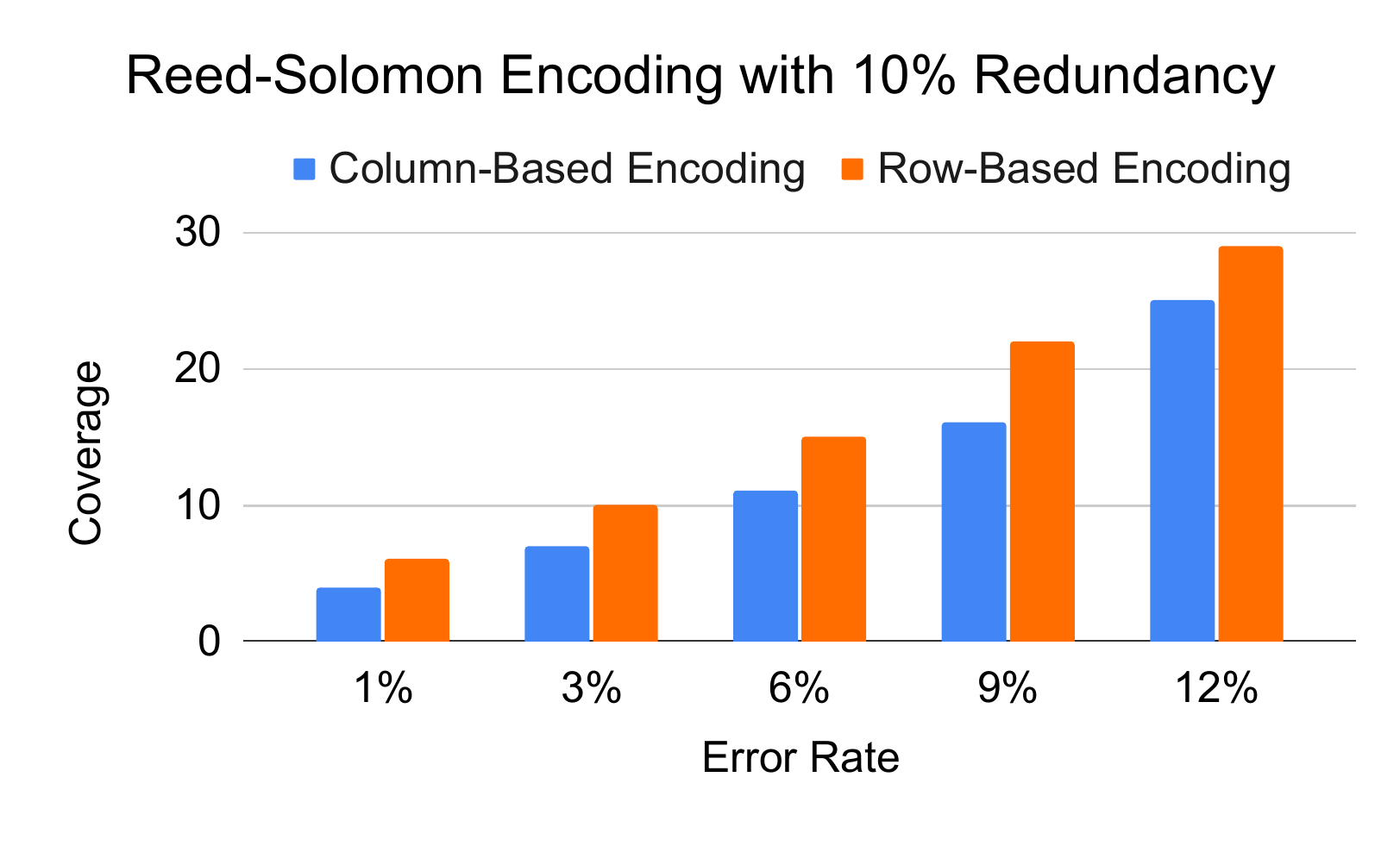}

    \caption{Minimum coverage required by our column-based and row-based implementations for decoding a 3MB archive file encoded with RS code at 10\% redundancy. Lower is better.}
    \label{fig:10p-comparison-reed-solomon}
\endminipage\hfill
\end{figure}

Our work is orthogonal to current efforts in designing optimal codes. Our core contributions include the columnar layout, integrated consensus, and block-based random access, all of which can be applied to any error-control codes. To demonstrate this, as we mentioned earlier, we have also implemented RS code with the same block length and symbol size  as used in Organick et al.~\cite{organick2018random}(65,536 symbols with 16 bits per symbol) in both columnar-based and row-based encoding implementations of CMOSS.
For this experiment, we encoded the same 3MB TPC-H archive file, using the same block length as Organick et al.\cite{organick2018random}, while maintaining an oligo length of 800nt. As a result, we generated 34,951 encoding oligos with the redundancy for RS code configured to 10\%. We used our simulator to vary the error rate between 1\% and 12\% similar to the LDPC experiment. For each error rate, and for each of the two layouts (column/row based), we generated read datasets at various coverage levels (from 1× to 25×). As shown in Figure~\ref{fig:10p-comparison-reed-solomon}, the trend of minimum coverage for various error rates is similar to the experiment conducted using LDPC; column-based encoding with its integrated consensus outperforms the row-based implementation even for RS as it requires lower coverage to fully decode the input data for all the error rates simulated. This shows that the columnar layout and integrated consensus aspects of CMOSS design are orthogonal to the choice of error-control codes.

Finally, we conclude this section by presenting a performance evaluation of our columnar implementation compared to the row-based one. We measured the run time for the experiments presented in Figure~\ref{fig:30p-coverage}. By varying the error rates and therefore the coverage required to reconstruct the data, the columnar-layout runtime varies between 35-39 minutes; the row-based implementation on the other hand takes around 5-6 minutes. The difference in time is due to the fact that in the columnar version, every LDPC block stored in  columns is re-encoded during the decoding process as shown in Figure~\ref{fig:dsm-rd-wt-pipeline} in steps (11)-(12). Moreover the freshly generated column of motifs is aligned against reads to fix the starting point of the next column of motifs (Figure~\ref{fig:dsm-rd-wt-pipeline}, steps (12)-(13)). 
Given the same number of oligos and the same coverage for both row-based and column-based encoding these two overheads will add a constant factor to the decoding time of column-based encoding compared to the row-based encoding for each OB that is being decoded. However this is not a scalability problem as (i) sequencing itself will take much longer than decoding, (ii) the decoding of OBs can be trivially parallelized, and (iii) in the context of long-term storage, decoding will be done once after several years or decades, and hence is the performance of decoding is not a limiting factor.


\subsection{Simulation of Large Scale Random Access}
\label{sec:larg-scale}

This section delves into the performance of our system in the accurate retrieval of only a small portion a large 100GB file. We encoded this file using LDPC codes with a redundancy rate of 30\%. Then, we mapped the resulting bits into oligos of 1024 nucleotides each. Based on these parameters, every OB is approximately 4MB in size, resulting in about 26,421 total OB. Every OB comprises 22,188 oligos, and we assume for this experiment that the number of OE matches the number of OB. In this way, a random access to one OE will correspond to a random access one OB, or 4MB. Every OE is then tagged with unique left and right primer sequences, extending the total nucleotide count to 1064 per oligo. The entire file is encoded using nearly 586M oligos.

To mimic real-world conditions, we developed a sampling algorithm that generated simulated reads based on the total oligos dataset and corresponding primers of the target extent. The algorithm included a similar rate of improper binding as found in earlier wet-lab experiments. Further, we used the BBMap tool to introduce random nucleotide variations, mimicking natural errors like substitutions, insertions, and deletions that characterize every sequencing process.

We then generated read datasets of various sizes to identify the minimum coverage required for accurate OE retrieval. We found that at least 155,052 reads were needed for complete reconstruction, assuming each oligo appeared seven times. However, nearly 26\% of the simulated reads contained errors—either in the primers or the payload—making them unsuitable for data retrieval. The remaining 74\% of reads were successfully used to retrieve the target extent. Subsequent analysis revealed that this corresponds to a 5x coverage rate, which we confirmed as the minimum necessary for accurate extent retrieval.

This experiment thus validates the efficacy of CMOSS in isolating and decoding specific OB within large files.

\section{Conclusion}

All SOTA approaches for DNA data archival use an object-based inteface and a horizon layout approach for mapping input bits onto oligos. In this paper, we showed how these assumptions amplify PCR coverage bias under a complex pool with files of multiple sizes, and lead to a strict separation of consensus calling and decoding, which, in turn, results in lost opportunity for improving read/write cost. We presented CMOSS, an end-to-end pipeline for DNA data archival that uses a novel, vertical oligo layout based on motifs as building blocks, and a fixed-size, block/extent-based random access over DNA storage. We showed how such an approach enables the merging of consensus and decoding stages so that errors fixed by decoding can improve consensus and vice versa. Using a full system evaluation, we highlighted the benefit of our design and showed that CMOSS can substantially reduce read-write costs compared to SOTA approaches.

\balance

\bibliographystyle{IEEEtran}
\bibliography{sample.bib}{}


\vfill

\end{document}


\maketitle





\noindent \textbf{Supplementary Figure 1.} Error rate per position in \textbf{\labone}.

\begin{figure}[htbp]
    \centering
    \includegraphics[width=0.6\textwidth]{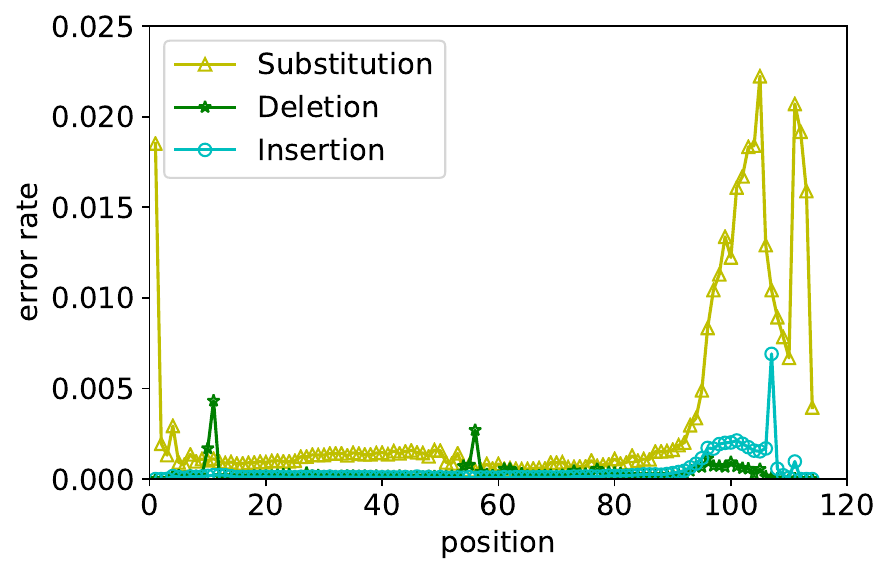}
    \label{fig:2-error-per-pos-3db-8tbl}
\end{figure}

\noindent \textbf{Supplementary Figure 2.} Error rate per position in \textbf{\labtwo}.

\begin{figure}[htbp]
    \centering
    \includegraphics[width=0.6\textwidth]{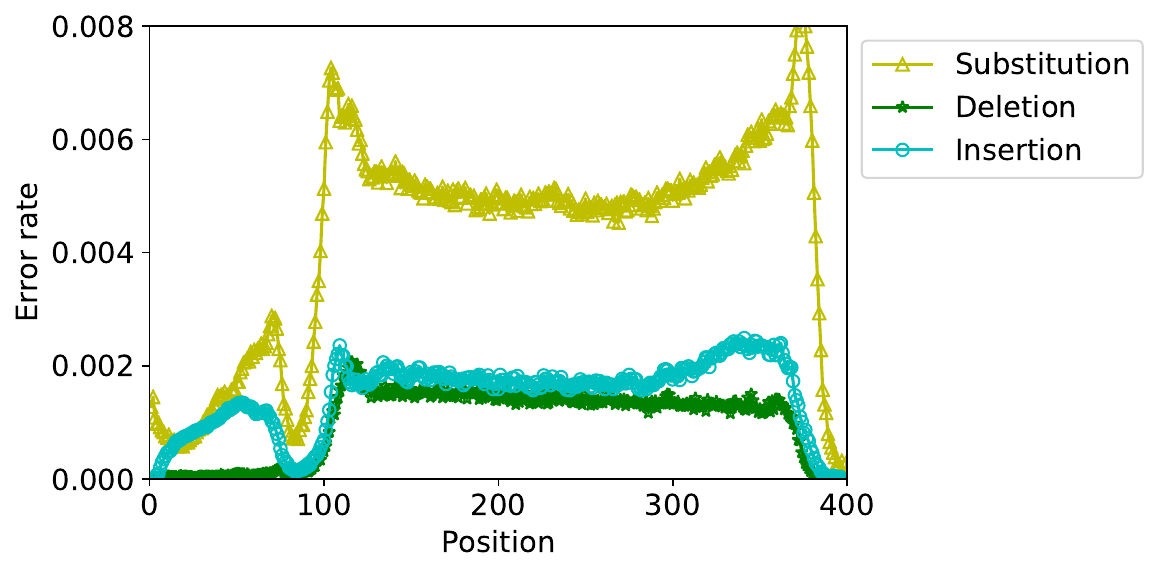}
    \label{fig:2-error-per-pos-4l-4r}
\end{figure}



\noindent \textbf{Supplementary Figure 4.} The oligo structure for object based abstraction.

\begin{figure}[htbp]
    \centering
    \includegraphics[width=0.5\textwidth]{IEEE-Transactions-LaTeX2e-templates-and-instructions/figure/3-lab1-structure.png}
    \label{fig:3-lab1-structure}
\end{figure}

\noindent \textbf{Supplementary Figure 5.} Histogram of coverage across oligos in \textbf{\labtwo}.
\begin{figure}[htbp]
    \centering
    \includegraphics[width=0.5\textwidth]{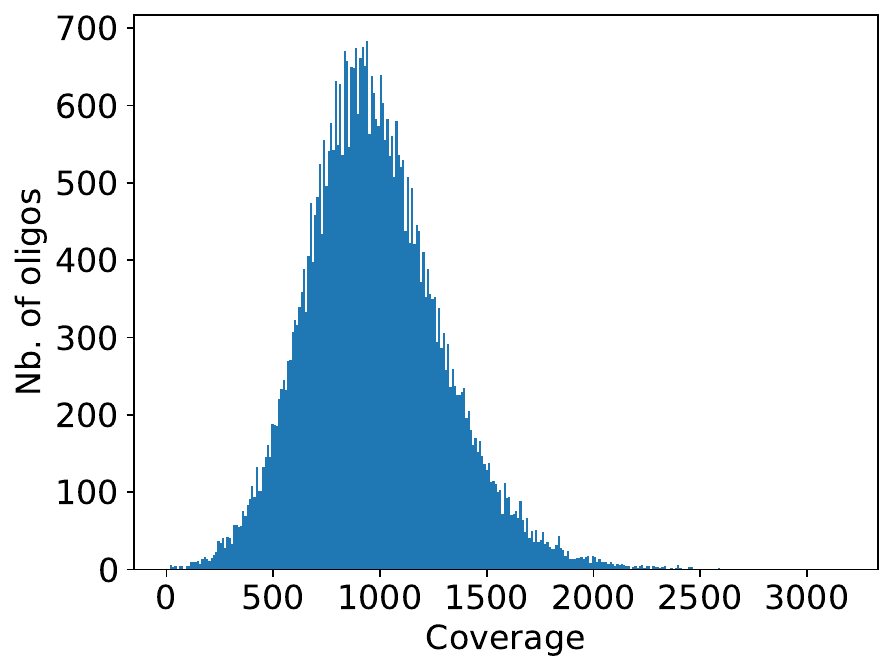}
    \label{fig:ref-his}
\end{figure}
    
\noindent \textbf{Supplementary Table 1.} The selected four left and four right primers used in \textbf{\labtwo} and \textbf{\labthree}.

\begin{table}[!htpb]
\centering
{
\begin{tabular}{rrrr}
\hline & left primers & & right primers \\\hline
L0 & ACATGCCGTGCCATTGGATT & R0 & AAGGCCAATTCGCGGTTAGT \\
L1& AGCCGACAAGTTCCAACACA & R1& AGGTGAGTGCCGTAACGATT \\
L2& GTCCAGGCAAAGATACAGTC & R2& GAACGGAGCGATGAGTTTGT \\
L3& TAGCCTCCAGAATGAGACAG & R3& TTCAAGCCAGACCGTGTGTA \\
\hline
\end{tabular}
}
\end{table}
